\documentclass[
reprint, twocolumn, secnumarabic%
,tightenlines%
,amssymb,
amsmath,
nobibnotes,
nofootinbib, aps, prl, showpacs,showkeys]{revtex4}

\usepackage{subfigure}
\usepackage{tikz}
\usepackage{docs}%
\usepackage{bm}%
\usepackage{graphicx} %
\usepackage{rotating} %
\usepackage{hyperref} %
\usepackage{url} %
\usepackage{srcltx} %
\expandafter \ifx \csname package@font\endcsname\relax\else
\expandafter \expandafter \expandafter
\usepackage
\expandafter \expandafter
\expandafter{\csname package@font\endcsname}%
\fi

\usepackage{epstopdf}
\usepackage{slashed}
\usepackage{ulem}
\usepackage{srcltx}
\usepackage{soul}
\usepackage{natbib} 
\usepackage{ulem} 

\newcommand{\bra}[1]{\langle #1 \rvert}
\newcommand{\ket}[1]{\lvert #1 \rangle}

\pacs{13.15.+g, 13.60.Le}

\keywords{neutrino-nucleon interactions, single pion production, lepton polarization, nucleon polarizations, non-resonant background contribution}

\usepackage{epstopdf}

\usepackage{enumerate}

\begin{document}

\title{Polarization Transfer in Weak Pion Production off the Nucleon}

\author{Krzysztof M. Graczyk}
\email{krzysztof.graczyk@uwr.edu.pl}

\author{Beata E. Kowal}

\affiliation{Institute of
Theoretical Physics, University of Wroc\l aw, pl. M. Borna 9,
50-204, Wroc\l aw, Poland}

\date{\today}%

\begin{abstract}
Polarization transfer (PT) observables in the single pion production  induced by the charged current interaction of the neutrino with the nucleon are examined. The polarization components of the final nucleon and the charged lepton are calculated within two  models for the pion production. The predictions are made for neutrino energy of the order of 1~GeV as well as for the T2K energy distribution. It is demonstrated  that the PT observables, the degree of polarization and the polarization components of outgoing fermions,  are  sensitive on  assumptions about the nonresonant background model. In particular it is shown that the normal components of the polarization of the outgoing nucleon and the lepton  are determined by the interference between the resonant (RES) and nonresonant (NB) amplitudes. Moreover the sign of the normal component of the polarization of the charged lepton is fixed by the relative sign between the RES and the NB amplitudes. 
\end{abstract}

\maketitle

\section{Introduction}

The first theoretical studies  of the single pion production (SPP) in neutrino-nucleon ($\nu N$) scattering were  performed more than forty years ago. A historical review of the topic can be found in \cite{LlewellynSmith:1971uhs}. Latest development of the accelerator neutrino oscillations experiments \cite{AguilarArevalo:2007it,Abe:2011ks,Evans:2013pka,Ayres:2004js} has triggered off  new interests in  the SPP processes. 

A further progress in the investigation of the fundamental properties of neutrinos: the oscillation phenomenon, the $CP$-violation in lepton sector, the mass hierarchy, requires an improvement of  the experimental techniques for the measurement of the interactions of neutrinos with different nuclear targets	\cite{Aliaga:2013uqz}  as well as  a development of the theoretical models  describing  the neutrino-nucleus scattering~\cite{doi:10.1146/annurev-nucl-102115-044720}. 

The measurement of the neutrino oscillation parameters  as well as extraction of the $CP$ violation phase are made based on the analysis of the quasielastic (QE) neutrino-nucleus scattering, however, the SPP processes contribute to the background for the detection of the QE-like events as well as  the electron neutrinos in the far detector. Hence the SPP contribution can not be neglected  in the analyses of neutrino scattering data. Additionally the investigation of the SPP in the $\nu N$ interactions allows to study the weak excitation of the nucleon to the resonance states. 

A first natural step in modeling the SPP in $\nu$-nucleus scattering is the construction of the theoretical description for the $\nu$-nucleon scattering. In this work we focus  on the  interactions of neutrinos with free nucleon target.  
Two mechanisms for the pion production in the $\nu N$ scattering can be distinguished: a resonant (RES) and a nonresonant. In the first the nucleon is excited to the resonance state, $N^*$, which subsequently decays to $\pi N$ system.  In the other there is no  $N\to N^*$ transition. This contribution can be modeled by  the so-called non-resonant background (NB) amplitudes allowed by the symmetries~\cite{Hernandez:2007qq}. 

The choice of the degrees of freedom  of the SPP model depends on the energy range in which it is applicable. If neutrino energy is relatively low, $E\sim 1$~GeV, it is enough to consider the contribution to the scattering amplitude from the first resonance region. For larger energies one has to include also the resonances states from the second and the third resonance regions  as well as higher order the NB terms. In this work we discuss neutrino energy of the order of 1~GeV, which is a kinematic domain typical  for the  long and short baseline  experiments with accelerator source of neutrinos \cite{AguilarArevalo:2007it,Abe:2011ks}. 


There are many phenomenological models describing the SPP in $\nu N$ interactions  \cite{Adler:1968tw,Rein:1980wg,Fogli:1979cz,Rein:1987cb,Hernandez:2006yg,Nakamura:2015rta,Serot:2012rd,Lalakulich:2010ss,Leitner:2008ue,Alam:2015gaa,Barbero:2008zza,Gonzalez-Jimenez:2016qqq}, for more complete list see \cite{Alvarez-Ruso:2014bla,doi:10.1146/annurev-nucl-102115-044720}.
One of the main difficulties in modeling the pion production is proposing a consistent description of both the RES and the NB contributions. The analysis of the unpolarized  cross section data does not allow to distinguish between the RES and the NB contributions. Moreover the main information about the SPP in the $\nu N$ scattering is obtained from the analysis of the data collected by two bubble chamber experiments ANL \cite{Radecky:1981fn} and BNL \cite{Kitagaki:1986ct} in which the neutrino-deuteron scattering was investigated. There are new measurements of the SPP in the $\nu$-nucleus scattering e.g. by Minerva~\cite{Altinok:2017xua} experiment. But in the analysis of these data the nuclear structure effects must be included which makes  the studies complicated and  model-dependent. 

In this work we propose to study the polarization properties of the particles in the SPP processes.   We shall show that the polarization transfer (PT) observables contain the non-trivial information about the resonance and the nonresonace transition. In particular we demonstrate that the investigation of  the normal polarization components  of the charged lepton and the final nucleon  give a knowledge about the relations between  the RES and the NB contributions. Indeed
the normal polarizations  are proportional to interference of the RES and NB amplitudes.  

The PT observables have been studied experimentally and theoretically in electron scattering off the nucleon/nuclei for last years. More than forty years ago Akhiezer et al. \cite{Akhiezer:1968ek,Akhiezer:1974em} (see also \cite{Ohlsen:1972zz,Arnold:1980zj,Donnelly:1985ry}) showed that from the measurement of the PT observables in the elastic  electron scattering off the proton the form factor ratio $G_E^p/G_M^p$ ($G_{E/M}^p$ is electric/magnetic proton form factor of the proton) can be obtained. This  is an alternative method for the measurement of the elastic form factors to the famous Rosenbluth method.  The ratio $G_E^p/G_M^p$  obtained from the PT data turned out to be inconsistent with the Rosenbluth measurements. It triggered off more detailed  experimental and theoretical studies of the elastic electron-nucleon scattering.  A recent review of the topic can be found in~\cite{Afanasev:2017gsk}.       

The investigation of the polarization properties of final particles in the $\nu N$ scattering is not a new idea. The PT observables in the QE $\nu N$ scattering were discussed in~\cite{LlewellynSmith:1971uhs}. Recently the problem has been refreshed in  works \cite{Bilenky:2013iua,Bilenky:2013fra,Akbar:2017qsf,Akbar:2016awk}. Moreover, in Refs. \cite{Kuzmin:2003ji,Hagiwara:2003di,Kuzmin:2004yb} the polarization properties of the $\tau$ lepton produced in the QE and inelastic $\nu N$ interactions were studied.  Additionally in \cite{Graczyk:2004uy} the impact of the  nuclear effects on the polarization of the $\tau$ lepton produced in the QE neutrino-nucleus  scattering was investigated.

The polarization properties of the $\tau$ lepton  in the SPP induced by the $\nu_\tau N$ scattering processes were studied by two groups:  Hagiwara \textit{et al.} \cite{Hagiwara:2003di} and Naumov \textit{et al.} \cite{Kuzmin:2003ji}.   However, the discussed SPP models did not contain the NB contribution.  In our work  the NB contribution  plays a central role. We show that the PT observables are sensitive on  the NB contribution.  We discuss the charged current interactions of the muon and the tau neutrinos  with the nucleons. Eventually we investigate also the polarization properties of the final nucleon produced in the SPP process. This problem  has been not studied before. 

We show that the PT observables are sensitive on the various details of the SPP models, in particular the description of the NB contribution. In order to study 
the model-dependence of the predictions of the PT observables we consider two phenomenological  approaches for the SPP: Hernandez-Nieves-Valverde (HNV) model  as formulated in~\cite{Hernandez:2007qq} and  Fogli-Nardulli (FN) model as described in~\cite{Fogli:1979cz}.  Both approaches are similar in the construction but it is demonstrated that small differences in the treatment of the NB contribution give rise to  disparities in the predictions of the PT observables. 
	
The paper is organized as follows: Section \ref{Section_Kinematics} introduces kinematics and the cross section formula, in Sec. \ref{Section_Polarization} the polarization observables are given,  Sec. \ref{Section_Models} contains short review of the HNV and the FN models, while in Sec. \ref{Section_Results} the  numerical results are presented and discussed. We summarized in Section \ref{Section_Summary}. Additionally  we include  three Appendixes.

\begin{figure}[t]
\includegraphics[width=0.5\textwidth]{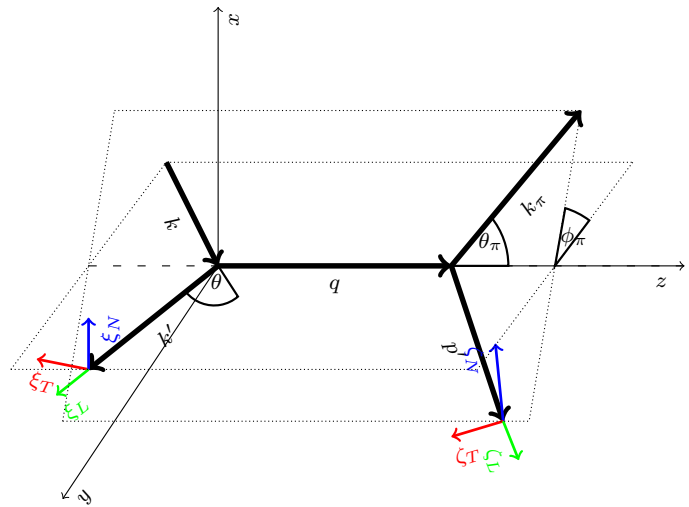}
	\caption{ \label{Fig_rysunek_przestrzeny_oddzialywania}  Angular distribution  of the particles, in the laboratory frame, in the process (\ref{Process_general}). The momenta $k$ and $k'$ denote the neutrino and the charged lepton respectively.  The target is at rest while $k_\pi$ and $p'$ denote the pion and the outgoing nucleon momenta. The vectors $\xi$ and $\zeta$ in green, blue and red denote longitudinal, normal, and transverse components of the charged lepton and the nucleon respectively.}
\end{figure}
\section{Kinematics And Cross Section}

\label{Section_Kinematics}

We consider the SPP induced by the charged current (CC) $\nu N$ interaction
\begin{equation}
\label{Process_general}
\nu_l(k) +N(p) \to  l^{-}(k') + N'(p')+\pi(k_{\pi}),
\end{equation}
where $l=\mu,\tau$, $k^\alpha=(E,{\bf k})$ and ${k'}^\alpha=(E',{\bf k'})$ are the four-momenta of the initial and the final leptons respectively, while $p^\alpha=(E_p,{\bf p})$, ${p'}^\alpha=(E_{p'},{\bf p'})$ and $k_\pi^\alpha=(E_{\pi},{\bf k_{\pi} })$ denote the four-momenta of the incoming nucleon (N), the outgoing nucleon ($N'$) and the pion respectively. Notice that 
$E_x=\sqrt{\mathbf{x}^2+M_x^2}$.
$M$, $m$ and $m_\pi$ denote  masses of  the nucleon, charged lepton and the pion respectively.

The four-momentum transfer is defined as
\begin{equation}
q^\alpha \equiv k^\alpha-{k'}^\alpha = (\omega,\mathbf{q}) 
\end{equation}
and the invariant hadronic mass $W$ is given by 
\begin{equation}
W^2 = (p+q) \cdot (p+q) \equiv (p+q)_\mu  (p+q)^\mu =(p+q)^2.
\end{equation} 
Let us also define 
\begin{equation}
Q^2 \equiv - q^2.
\end{equation}
By $\theta \equiv \angle(\mathbf{k},\mathbf{k'})$ the scattering angle between lepton momenta is denoted while by $\Omega$
the spherical angle (depending on $\theta$) is denoted;
 $\phi_\pi$ is the angle between the scattering plane (spanned by $\mathbf{k}$ and $\mathbf{k'}$) and the plane spanned by the pion and the final nucleon momenta, see Fig. \ref{Fig_rysunek_przestrzeny_oddzialywania}. 
 
\begin{figure}[t]
\subfigure[]
{
	\includegraphics[width=1.5in]{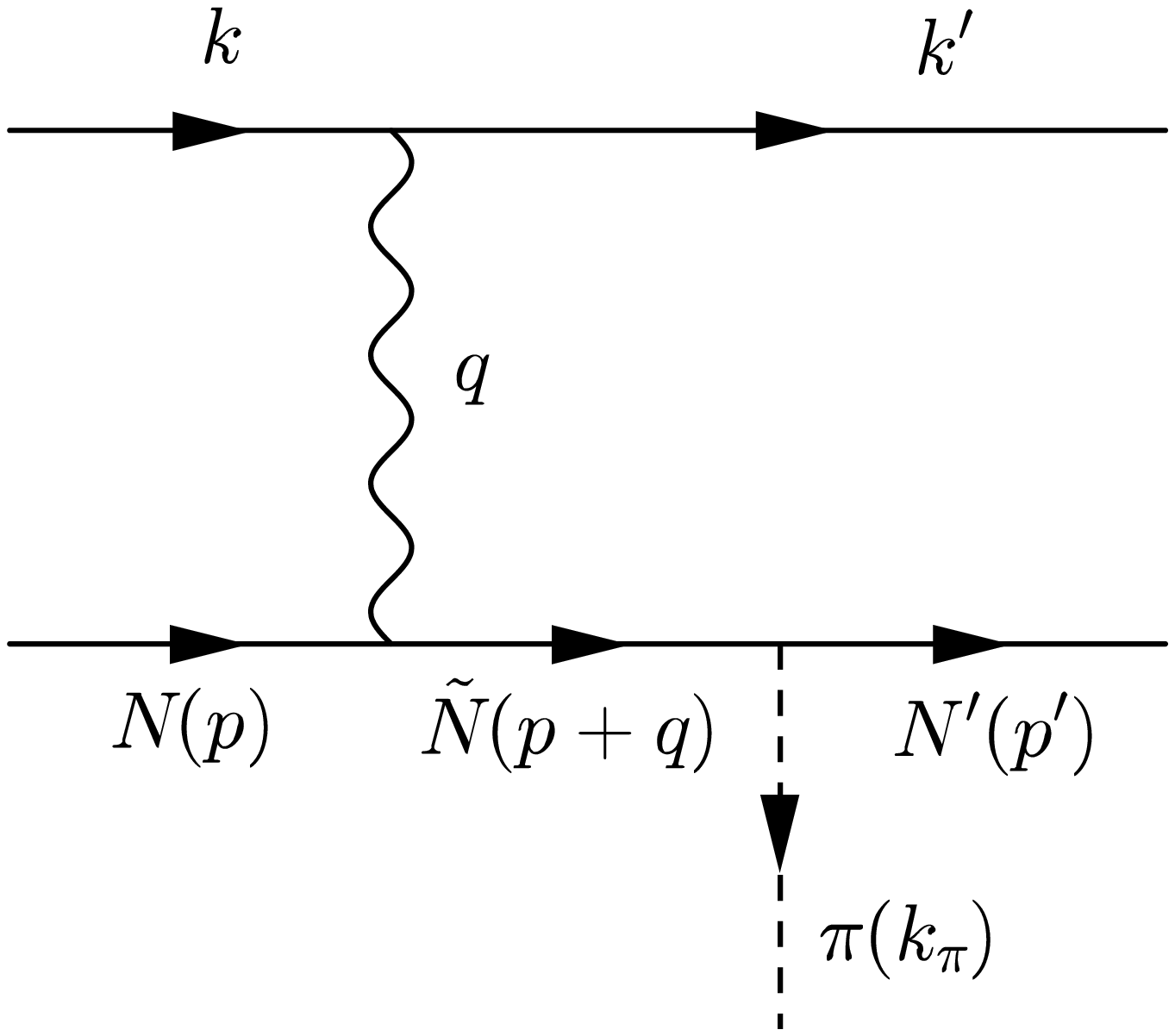}
}
\subfigure[]
{
	\includegraphics[width=1.5in]{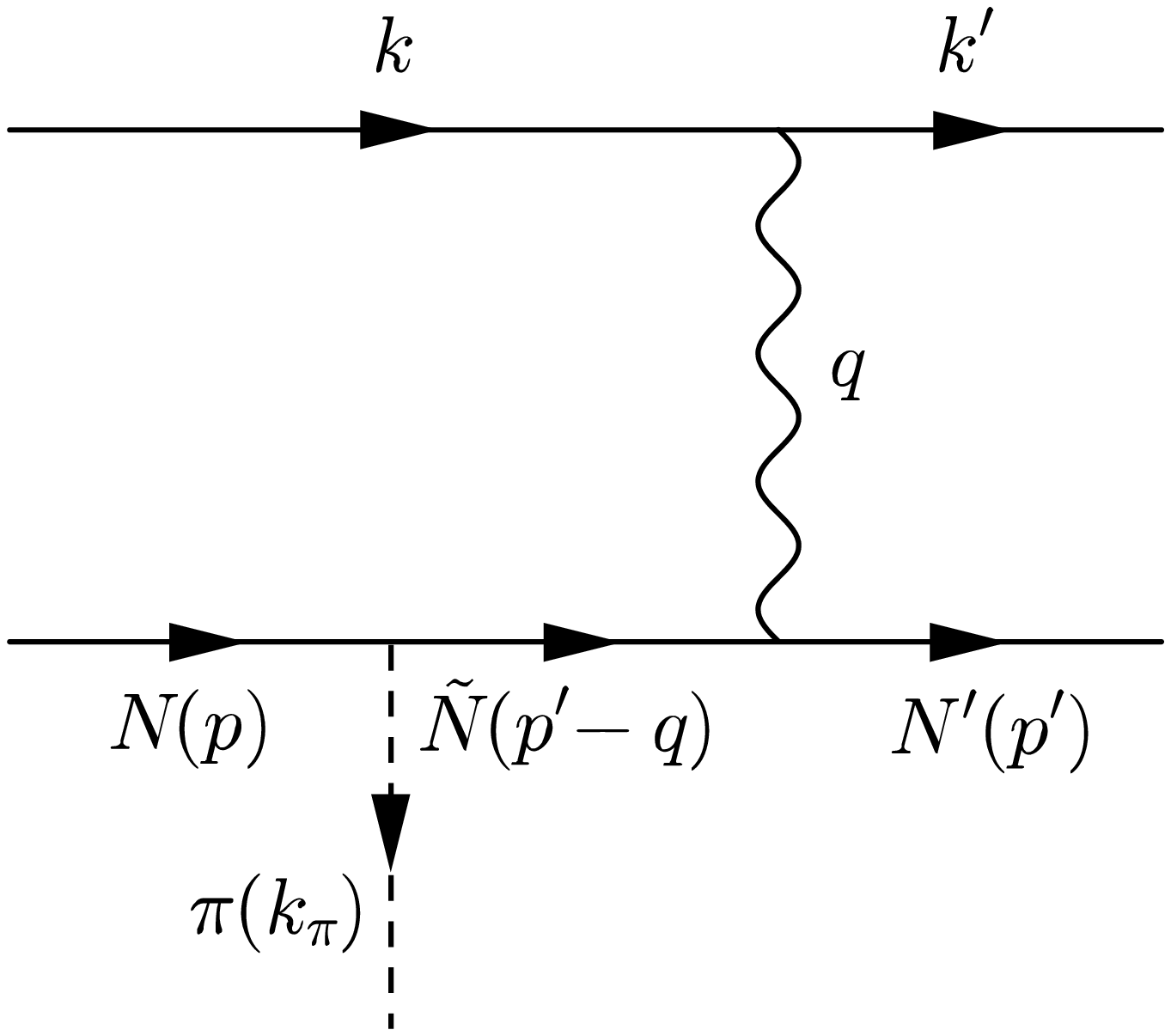}
}
\\
\subfigure[]
{
	\includegraphics[width=1.5in]{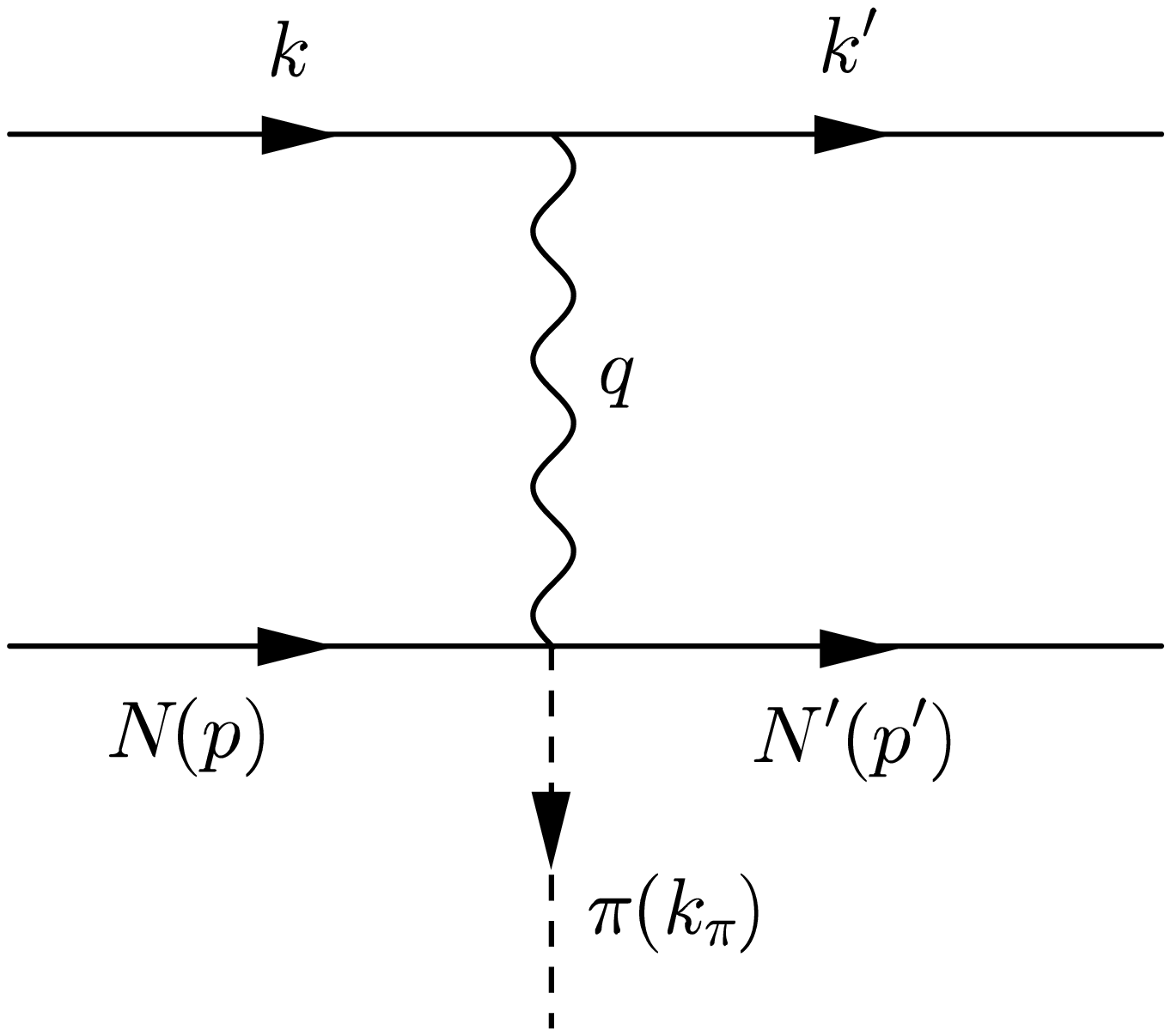}
}
\subfigure[]
{
	\includegraphics[width=1.5in]{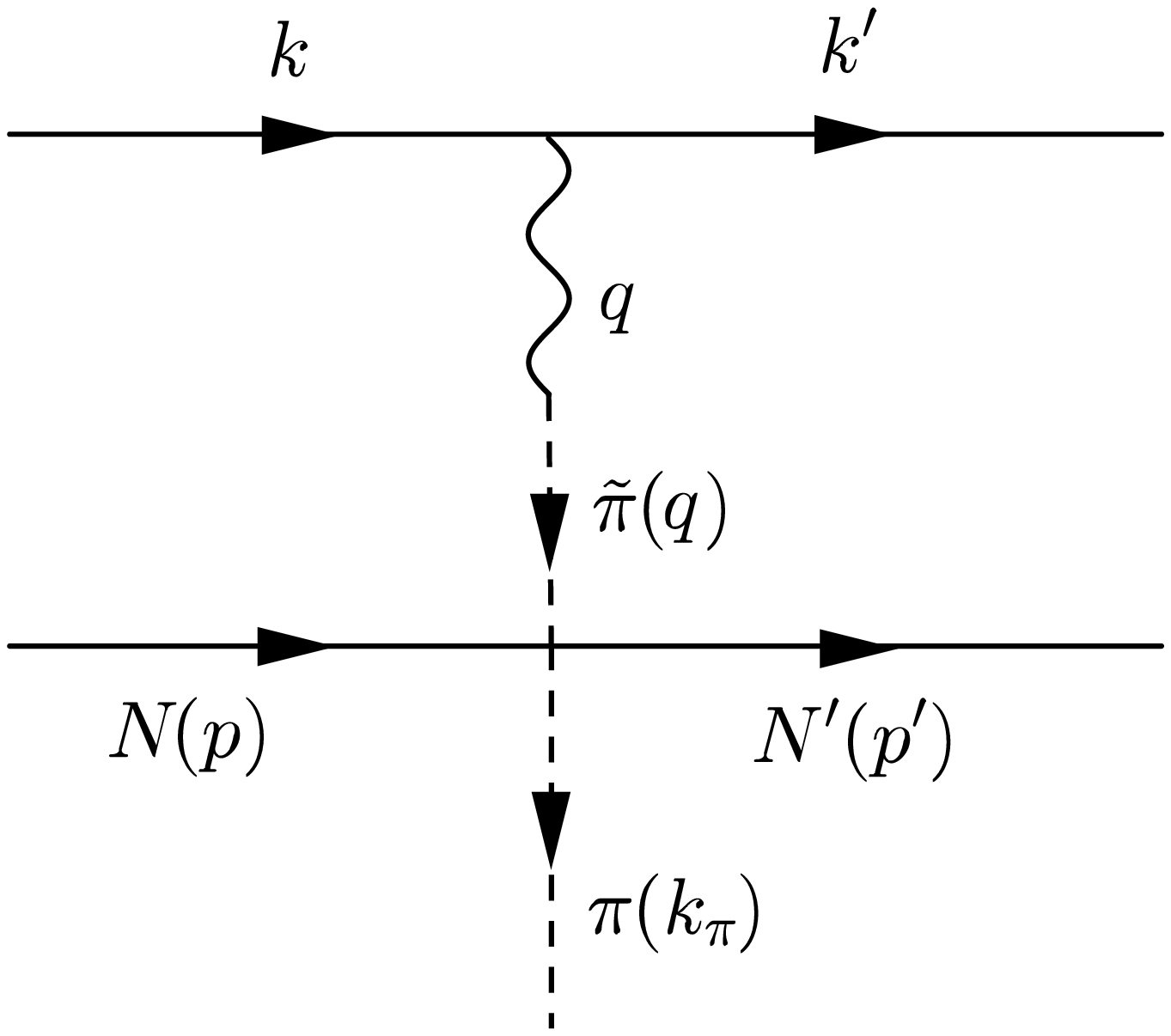}
}
\\
\centering
\subfigure[]
{
	\includegraphics[width=1.5in]{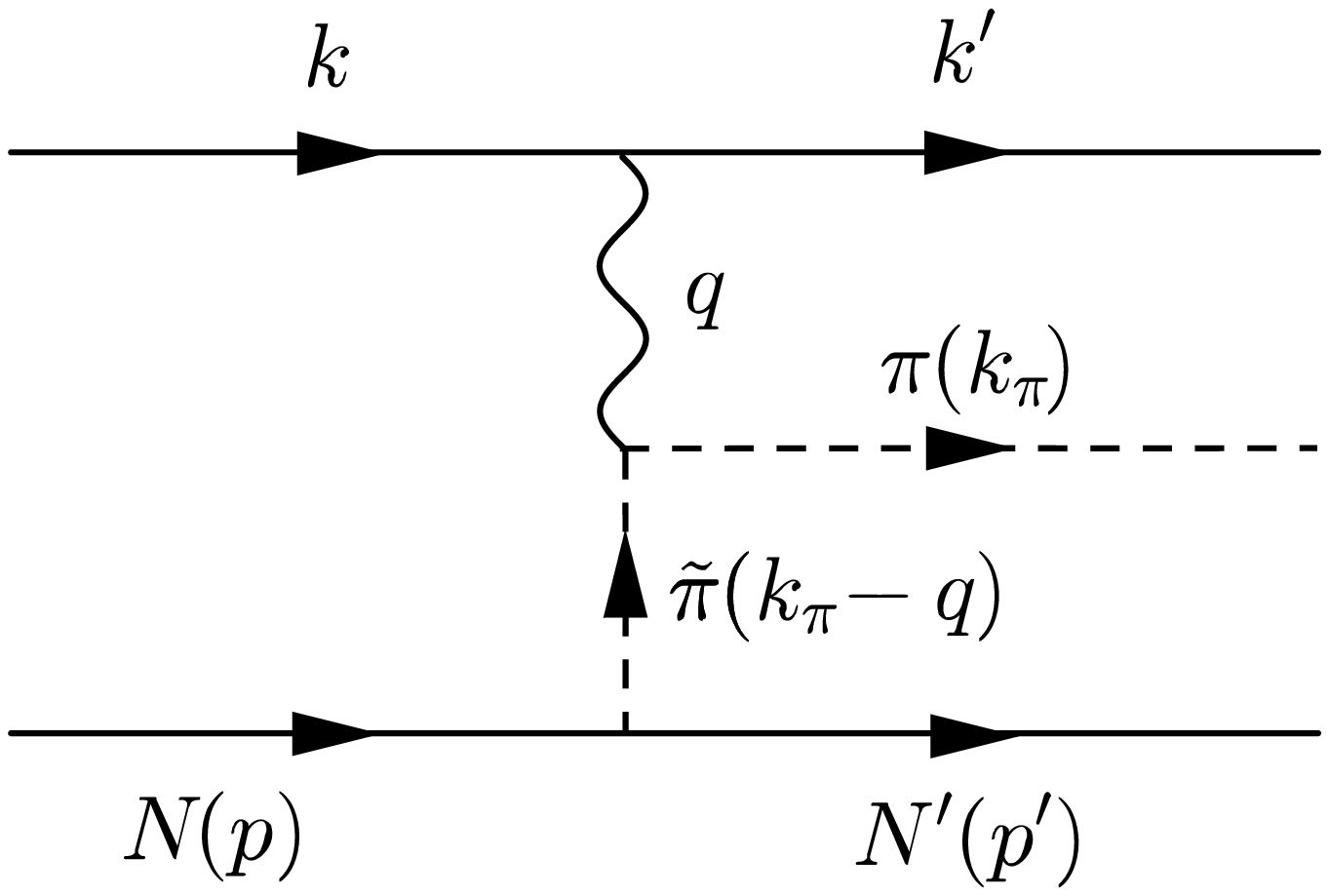}
}
\\
\subfigure[]
{
	\includegraphics[width=1.5in]{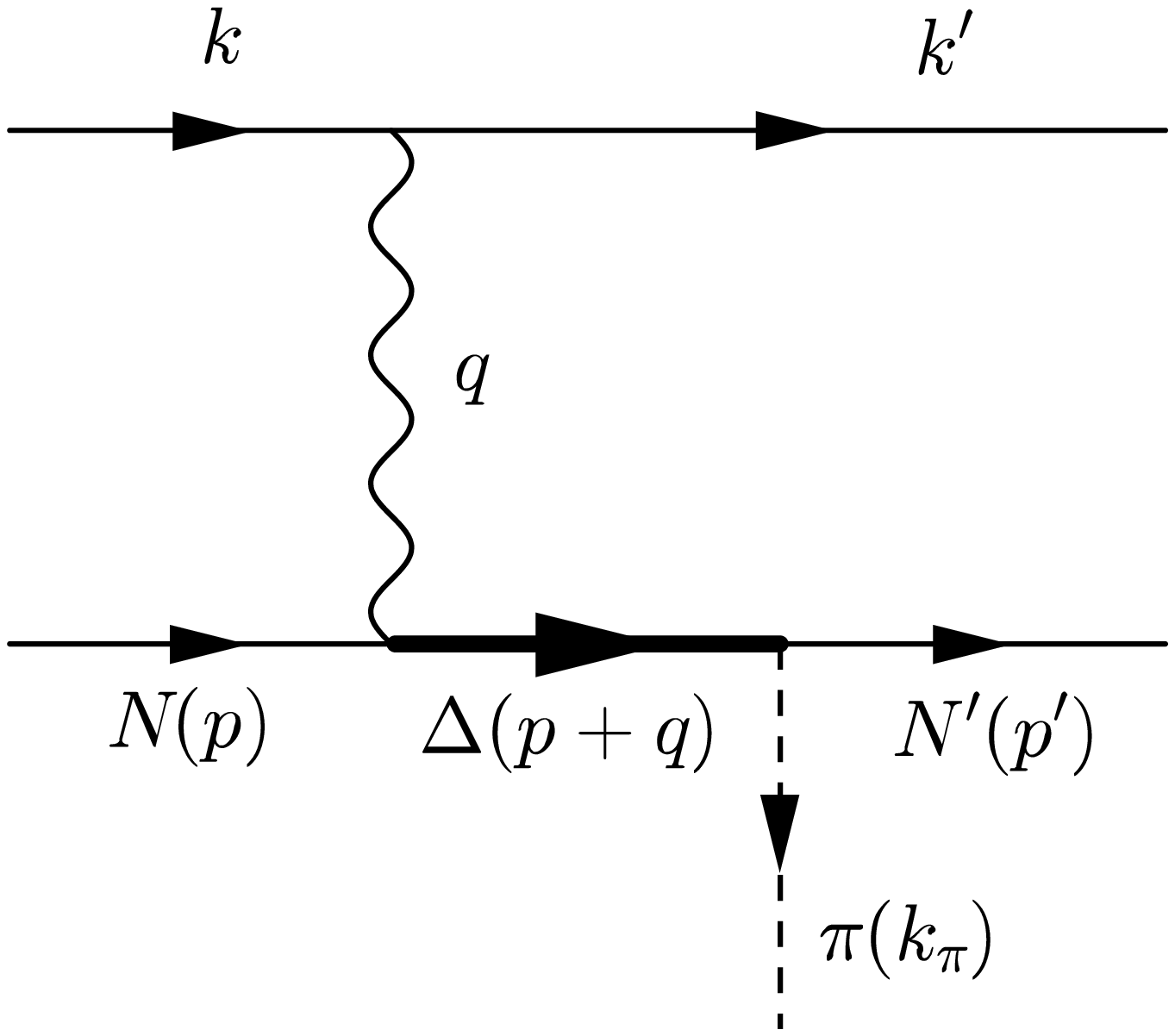}
}
\subfigure[]
{
	\includegraphics[width=1.5in]{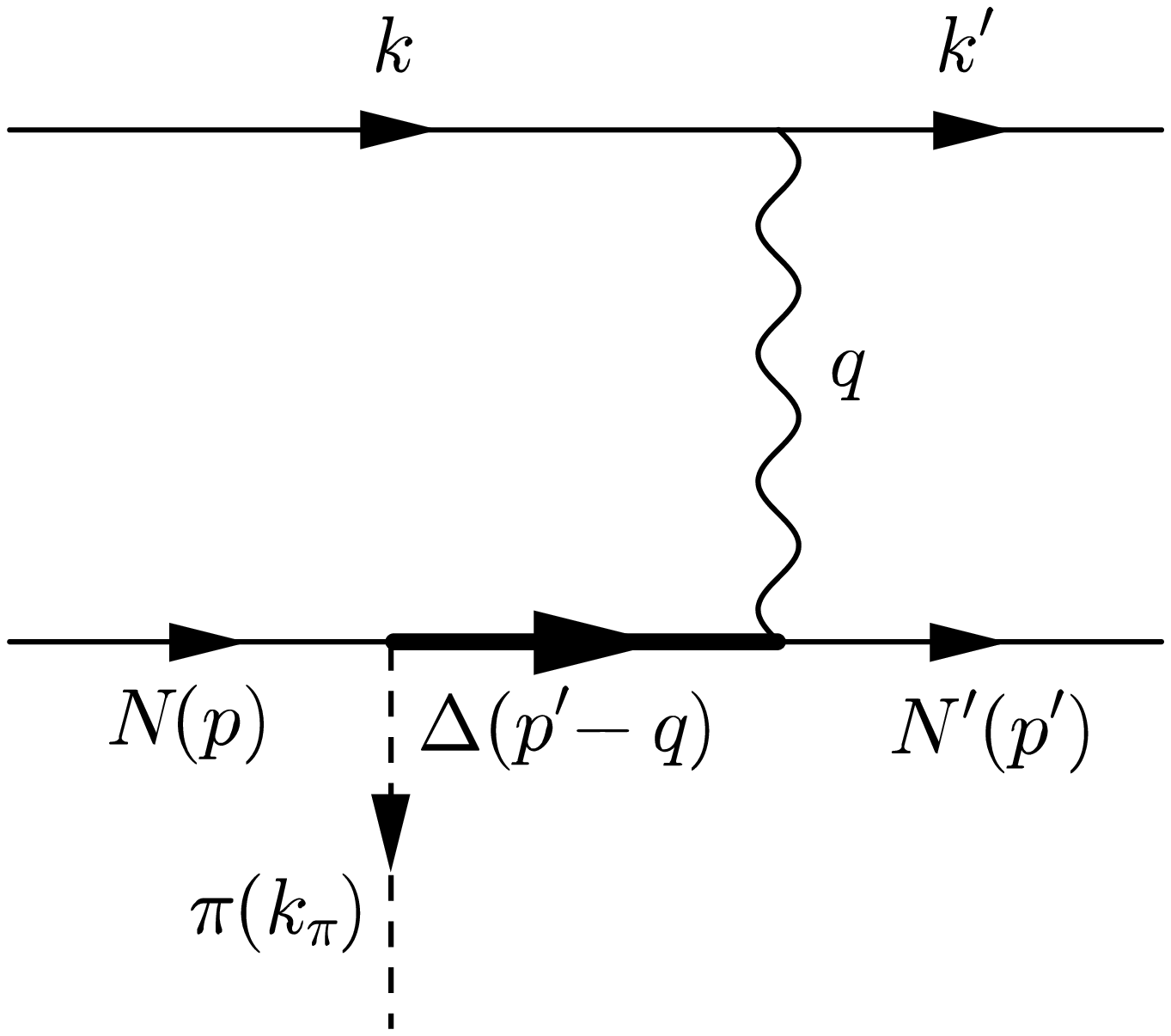}
}
\caption{Diagrams for the SPP in $\nu N$ scattering. The NB contribution is given by: a) nucleon pole (NP); b) crossed nucleon pole (CNP); c) contact term (CT); d) pion pole (PP); e) pion in flight (PF). The RES contribution is given by: f) delta pole ($\Delta$P); g) crossed delta pole (C$\Delta$P). $\tilde{N}$ denotes a virtual nucleon. 
\label{Fig_diagrams}} 
\end{figure}

The differential cross-section for the process (\ref{Process_general}) reads:
	\begin{eqnarray}
	d\sigma & = & \frac{1}{4 M E_{k}}  
	\sum_{spins}
	\;{\sum_{spins}}'
	\frac{d^3 \textbf{k'}}{2 E_{k'} (2 \pi)^3}\frac{d^3 \textbf{p'}}{2 E_{p'} (2 \pi)^3}\frac{d^3\textbf{k}_{\pi}}{2 E_{\pi} (2 \pi)^3}
	\nonumber \\
	& &\label{cross_section_general_formula}
	(2\pi)^4 \delta^{(4)} (p'+k_{\pi}+k'-k-p) |\mathcal{M}_{fi}|^2.
	\end{eqnarray}
Sum over the spins of the leptons and the nucleons is denoted by  ${\sum}'$  and $\sum$ respectively. We systematically omit the spin notation in this section.  
\begin{figure*}
	\centering{
	\includegraphics[width=0.9\textwidth]{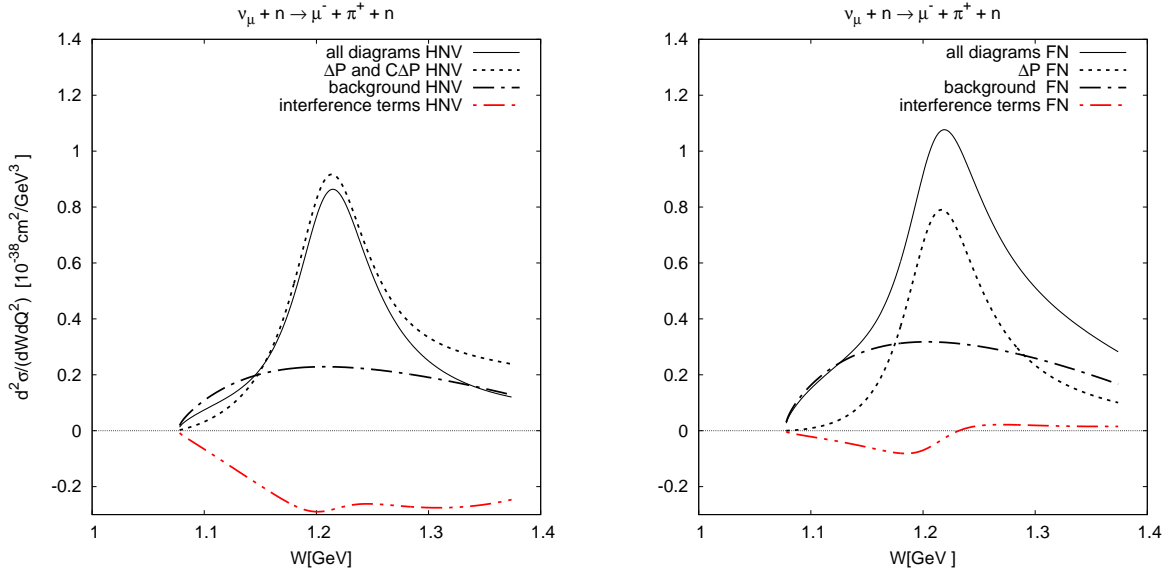}
		\caption{\label{Fig_partition_of_the_nonresonant}
			The differential cross section and its  ''pure'' resonance (dotted), ''pure'' nonresonant (dashed-dotted) and the interference between resonant and nonresonant amplitudes  (dashed-dotted-doted)  contributions in the HNV (left figure) and the FN (right figure) models. The "pure" resonance contribution is given by $|\mathcal{M}_{\Delta P} + \mathcal{M}_{C\Delta P}|^2$ and $|\mathcal{M}_{\Delta P}|^2$ in the HNV and the FN models respectively.  The neutrino energy $E=0.7$~GeV and $Q^2=0.1$~GeV$^2$.}
	}
\end{figure*}

The scattering matrix, in the one-boson exchange approximation and for $-q^2 \ll M_W^2 $ reads:
\begin{equation}
\mathcal{M}_{fi}
=
- \frac{G_F}{\sqrt{2}}   \cos\theta_C 
\bar{u}(k')\gamma_{\mu} (1-\gamma_5)  u(k)
\, \bra{ \pi , N' } \mathcal{J}^{\mu}  \ket{N},
\end{equation}
where $G_F\approx~1.16639\times~10^{-5}$~GeV$^{-2}$ is the Fermi constant  while $\theta_C$ is the Cabibbo angle and $ \cos\theta_C \approx 0.97427$.  
\begin{figure*}
	\centering{
	\includegraphics[width=\textwidth]{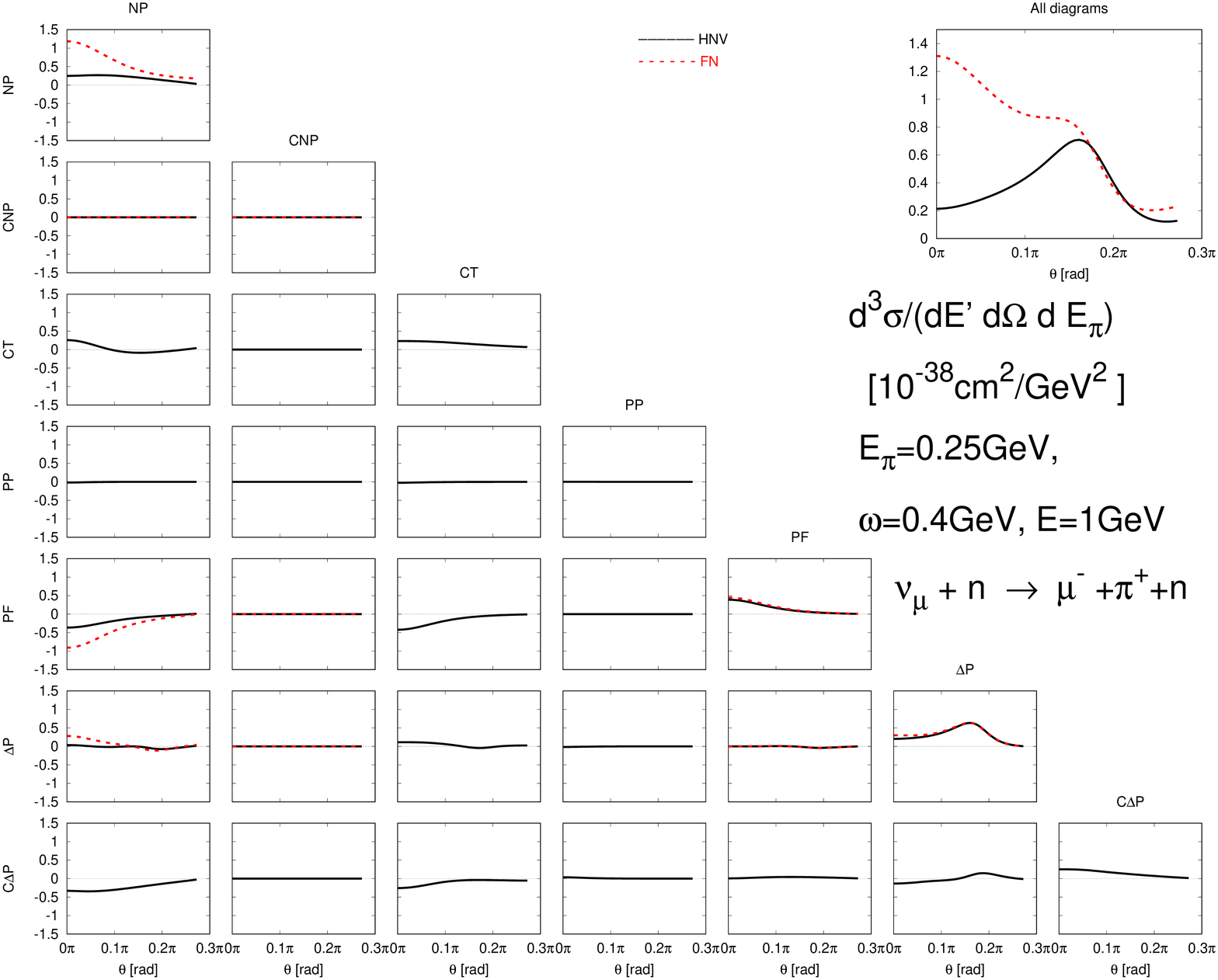}
     \caption{\label{Fig_cross_section_partition} 
     Separation of the differential cross section into various interference components.  The dotted/solid (black/red) line represents the HNV/FN model predictions.  The contribution from the $ |\mathcal{M}_a|^2 $  are on the diagonal while below the diagonal the interference terms  $2\Re (\mathcal{M}_i \mathcal{M}_j^*)$ ($j$ -- column, $i$ -- row) are plotted. 
     }
    }
\end{figure*}
The  expectation value of the CC weak hadronic current operator $\mathcal{J}^\mu$ reads
\begin{equation}
J^\mu  \equiv \bra{ \pi , N' } \mathcal{J}^{\mu}  \ket{N}.
\end{equation}

Distinct phenomenological models can contain different numbers of diagrams contributing to the total scattering amplitude. Hence the total hadronic current can be given by the sum:
\begin{equation}
J^\mu \to \sum_{a\in S} \mathcal{C}^a J^\mu_{a},
\end{equation}
where $\mathcal{C}^a$ is the Clebsch-Gordan coefficient, see Table \ref{Clebsch-Gordan-CC}.

The total amplitude in the HNV model is   described by  seven Feynman diagrams, hence
\begin{equation} 
\mathcal{S}_{HNV} = \{NP,  CNP,  CT, PP, PF, \Delta P, C\Delta P\}.
\end{equation} 
An explanation of the notation in the parentheses can be found in  Fig.~\ref{Fig_diagrams} and Ref.~\cite{Hernandez:2007qq}.

In the FN model the total amplitude is given by  four diagrams
\begin{equation} 
\mathcal{S}_{FN} = \{NP,  CNP, PF, \Delta P \}.
\end{equation}

The structure of the amplitudes in the HNV and FN models is the same, hence we introduce 
\begin{equation}
J^\mu_{a} = \langle \pi, N'\mid \mathcal{J}^\mu_a(0) \mid N\rangle = \overline{u}(p') R^\mu_a u(p),
\end{equation}
where $R_a^\mu = r^\mu_a$ and $R^\mu_a = \tilde r^\mu$ for the HNV and the FN models respectively.
 
Eventually it is convenient to introduce the notation 
\begin{equation}
\mathcal{M}_a = -\frac{G_F}{\sqrt{2}}\cos\theta_C  \bar{u}(k')\gamma_{\mu} (1-\gamma_5)  u(k) R^\mu_a,
\end{equation}
then 
\begin{eqnarray}
 \mathcal{M}_{fi}
&=&
\sum_{a \in \mathcal{S}} \mathcal{M}_a.
\end{eqnarray}

\section{Polarization Transfer Observables}
\label{Section_Polarization}

We consider two types of the PT processes. In the first the polarization of the final lepton is examined, namely
\begin{equation}
\label{Process_lepton_polarization}
\nu_l(k) + N(p)  \to   \vec{l}^{-}(k',\xi) + N'(p')+\pi(k_{\pi}).
\end{equation}
In the other the polarization of the final nucleon is a subject of studies
\begin{equation}
\label{Process_final_nucleon_polarization}
\nu_l(k) + N(p)  \to   l^{-}(k') + \vec{N'}(p',\zeta)+\pi(k_{\pi}).
\end{equation}
Notice that $\xi$ and $\zeta$ are the four-vector spins of  the final  lepton and the nucleon respectively.

A spin four-vector $s^\mu$ of a $1/2$-spin fermion of  mass $M$  has three independent components. At rest it has the form 
\begin{equation} 
s^\mu = (0,\hat{\mathbf{s}}), \;\;\; \mathrm{where}\,\;  \hat{\mathbf{s}}^2=1.
\end{equation}
In a frame in which  a particle moves with the velocity $\mathbf{p}/E_p$ the spin vector reads~\cite{Dombey:1969wk,Greiner__QED_book,Maximon:2000hk}:
\begin{equation}
s^\mu = \left( \frac{\boldsymbol{\hat{s}}\cdot \mathbf{p}}{M},\boldsymbol{\hat{s}} + \mathbf{p} \frac{\boldsymbol{\hat{s}}\cdot \mathbf{p}}{M(M+E_p)}\right), 
\end{equation}
At any frame 
\begin{equation}
s\cdot p  = 0.
\end{equation}
It is convenient to introduce: longitudinal ($L$), transverse ($T$) and normal ($N$) components of the spin
\begin{eqnarray}
\xi^\mu & = & \xi^\mu_L + \xi^\mu_T  + \xi^\mu_N,
\\
\zeta^\mu & = & \zeta^\mu_L + \zeta^\mu_T  + \zeta^\mu_N,
\end{eqnarray}  	
where we choose the coordinates so that (see Fig. \ref{Fig_rysunek_przestrzeny_oddzialywania})
\begin{eqnarray}
\xi^\mu_{L}
&=&
h_l \left(\frac{|\mathbf{k'}|}{m}, \frac{E_{k'}}{m}\frac{\mathbf{k'}}{|\mathbf{k'}|}\right),
\\
\xi^\mu_{N}
&=&  h_l \left(0, \frac{\mathbf{k} \times \mathbf{q}}{|\mathbf{k} \times \mathbf{q} |}\right), 
\\
\xi^\mu_{T}
&=&
h_l \left(0, \frac{\mathbf{k'} \times (\mathbf{k}\times \mathbf{q})}{|\mathbf{k'}\times (\mathbf{q}\times \mathbf{k})|}\right),
\end{eqnarray}
and $h_l = \pm 1$ as well as 
\begin{eqnarray}
\zeta^\mu_{L}
&=&
h_N \left(\frac{|\mathbf{p'}|}{M}, \frac{E_{p'}}{M}\frac{\mathbf{p'}}{|\mathbf{p'}|}\right),
\\
\zeta^\mu_{N}
&=&  h_N\left(0, \frac{\mathbf{p'} \times \mathbf{k}_\pi}{|\mathbf{p'} \times \mathbf{k}_\pi|}\right), 
\\
\zeta^\mu_{T}
&=&
h_N\left(0, \frac{\mathbf{p'} \times (\mathbf{p'} \times \mathbf{k}_\pi)}{|\mathbf{p'} \times (\mathbf{p'} \times \mathbf{k}_\pi)|}\right),
\end{eqnarray}
and $h_N = \pm 1$.

For the process (\ref{Process_lepton_polarization}) the differential cross section reads
\begin{equation}
d \sigma  \sim \frac{1}{2} \overline{\mid \mathcal{M}_{fi}\mid }^2 \left(1  + \mathcal{P}_l^\mu \xi_\mu \right),
\end{equation}
while for the process (\ref{Process_final_nucleon_polarization}) the cross section has the form: 
\begin{equation}
d \sigma  \sim \frac{1}{2} \overline{\mid \mathcal{M}_{fi}\mid}^2 \left(1 + \mathcal{P}_N^\mu \zeta_\mu \right),
\end{equation}
where $\overline{\mid \mathcal{M}_{fi}\mid }$ is summed over the spins.

The four-vectors $\mathcal{P}_\mu^l$ and $\mathcal{P}_\mu^N$ describe the  polarization of the charged lepton (l) and the final nucleon (N). The  components of polarization  are given by the ratio~\cite{Arnold:1980zj,Walecka_electron_scattering_textbook}: 
\begin{equation}
\label{Polarization_component_operational}
\mathcal{P}_{X}^Y = \frac{d \sigma( s^\mu_X)-d \sigma( -s^\mu_X)}{d \sigma( s^\mu_X) + d \sigma( -s^\mu_X)} = \mathcal{P}^\mu_Y s_{X\,\mu},
\end{equation}
where
$X=L$ (longitudinal), $N$ (normal) and $T$ (transverse) components of the polarization of the final fermion; $Y=l,N$ and 
$s=\xi$, $\zeta$.

Finally we define the degree of polarization of particle:
\begin{equation}
\label{degree_of_polarization}
\mathcal{P} = \sqrt{\mathcal{P}_L^2 + \mathcal{P}_N^2 + \mathcal{P}_T^2}.
\end{equation}
Notice that $ 0 <  \mathcal{P} < 1 $. If $\mathcal{P} \approx 0$ than  the particle is unpolarized, while for fully polarized particle $\mathcal{P} \approx 1$. 

\begin{figure*}
\centering{
\includegraphics[width=0.9\textwidth]{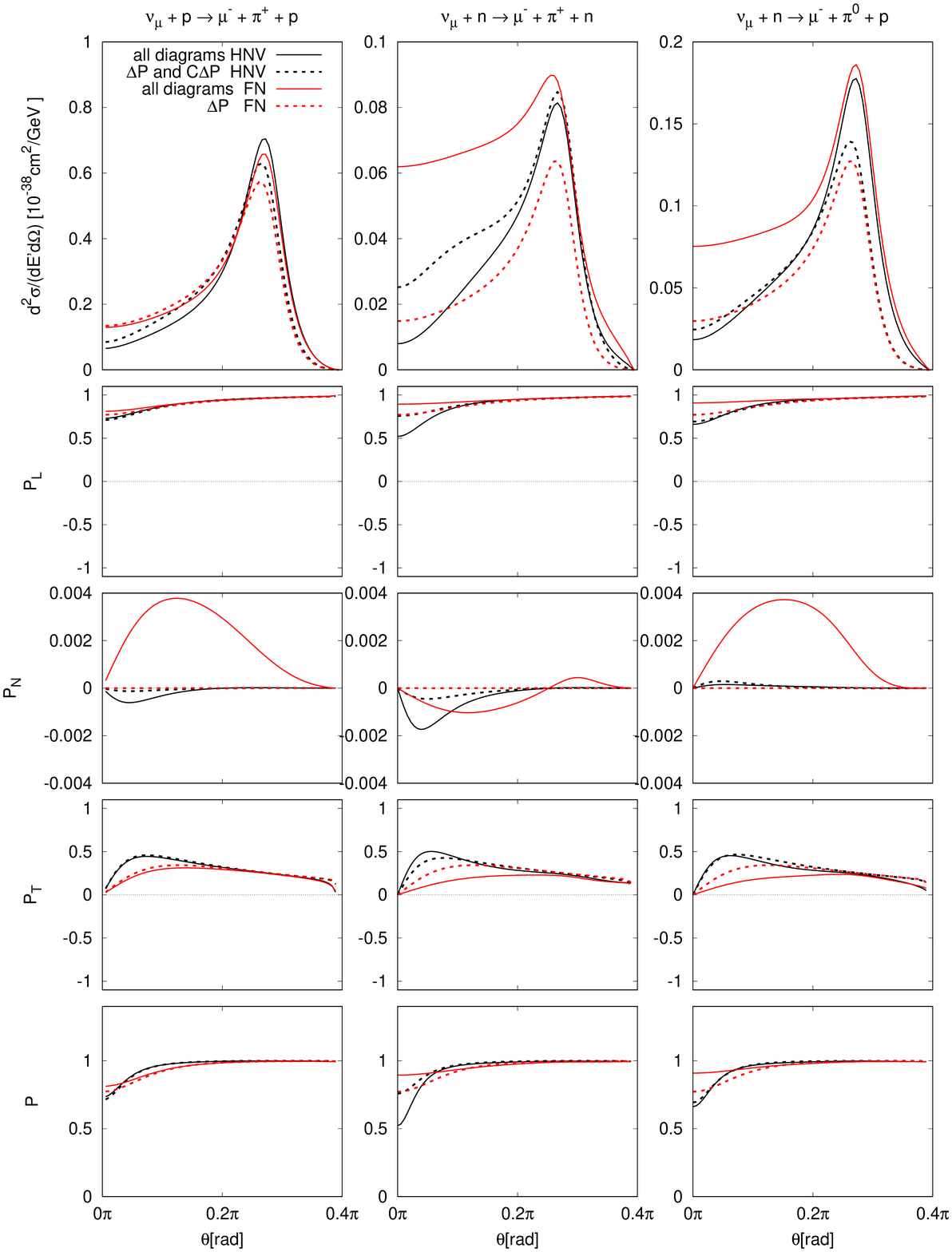}
\caption{\label{Fig_muon_neutrino_full_theta} 
Angular dependence of the components of  $\mathcal{P}_{\mu^-}^\alpha(E,\omega,\theta)$ vector polarization and the degree of polarization (last row) of the $\mu^-$ lepton, for the energy of the neutrino $E=1$~GeV and  the transfer  of energy $\omega=0.5$~GeV. The dotted/solid line represents the RES/full model contributions of the HNV (black) and   the FN (red) models. The resonance contribution is given by $|\mathcal{M}_{\Delta P} + \mathcal{M}_{C\Delta P}|^2$ and $|\mathcal{M}_{\Delta P}|^2$ in the HNV and the FN models respectively.
}
}
\end{figure*}
\begin{figure*}
\centering{
\includegraphics[width=0.9\textwidth]{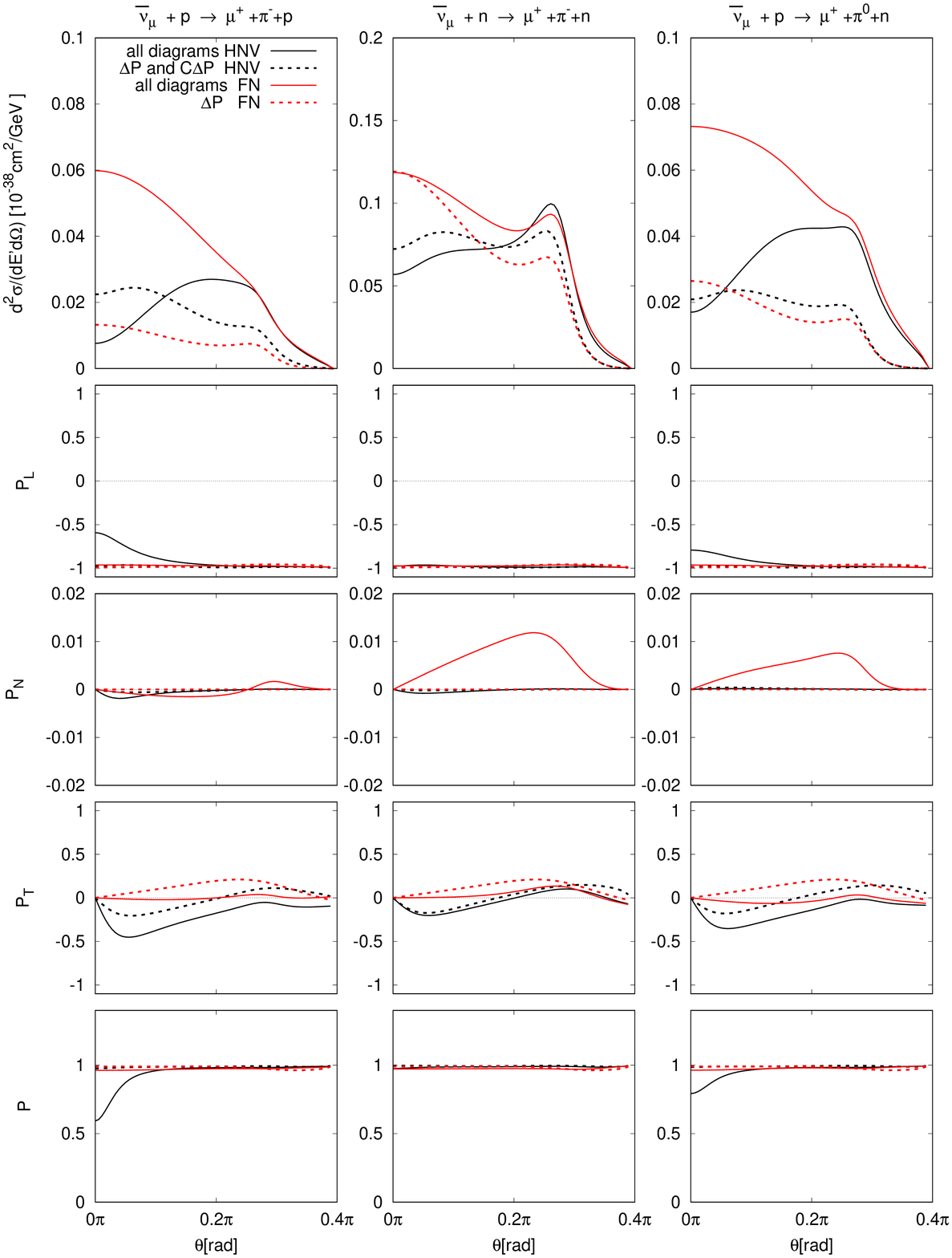}
	\caption{\label{Fig_muon_antyneutrino_full_theta} 
	Angular dependence of the components of  $\mathcal{P}_{\mu^+}^\alpha(E,\omega,\theta)$ vector polarization and the degree of polarization (last row) of the $\mu^+$ lepton, for the energy of the neutrino $E=1$~GeV and  the transfer  of energy $\omega=0.5$~GeV.
	The dotted/solid line represents the RES/full model contributions of the HNV (black) and   the FN (red) models. The resonance contribution is given by $|\mathcal{M}_{\Delta P} + \mathcal{M}_{C\Delta P}|^2$ and $|\mathcal{M}_{\Delta P}|^2$ in the HNV and the FN models respectively.
	}
	}
\end{figure*}
\begin{figure*}
\centering{
\includegraphics[width=0.9\textwidth]{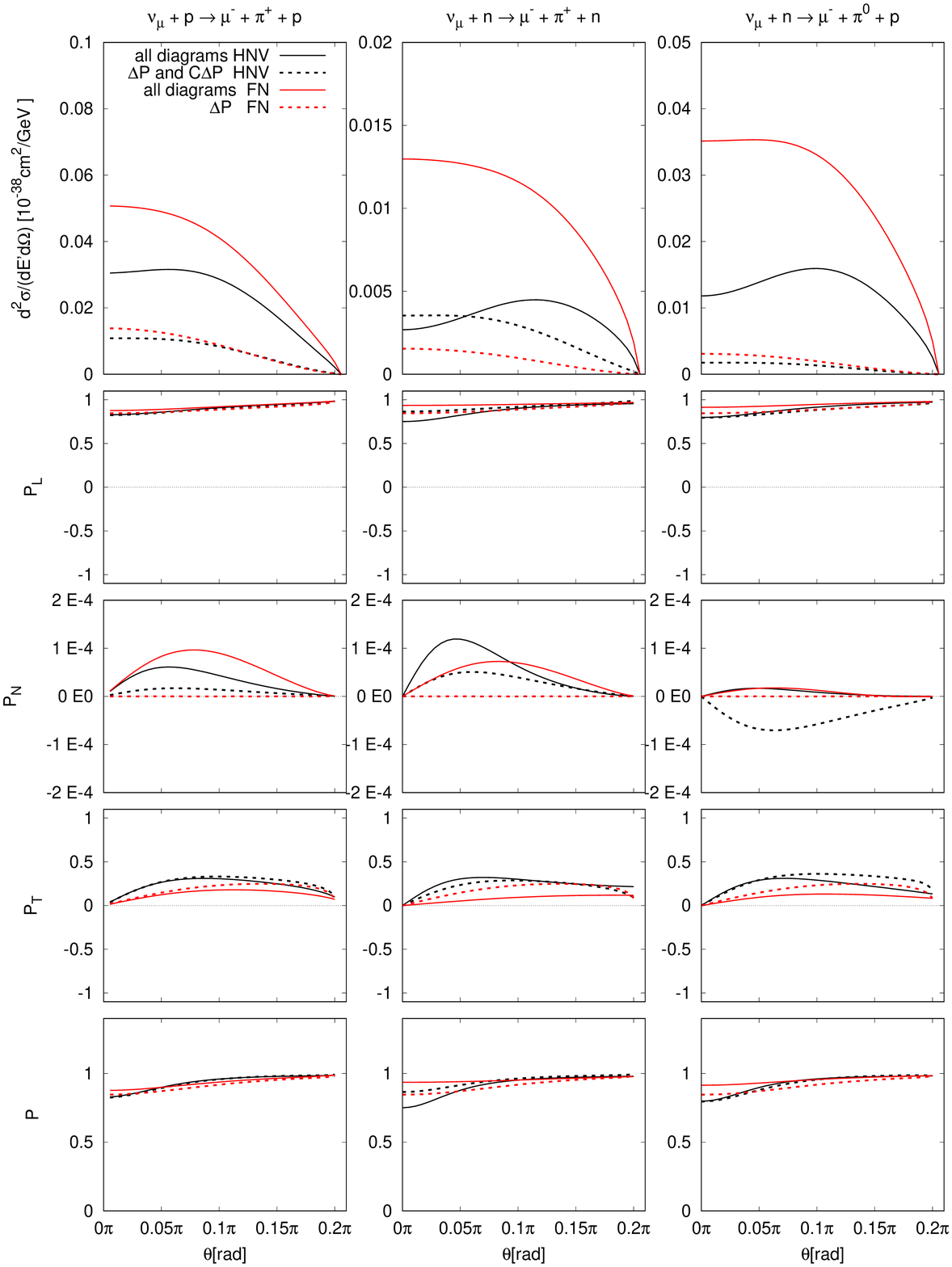}
\caption{\label{Fig_muon_neutrino_E=06_full_theta} 
Caption the same as in Fig. \ref{Fig_muon_neutrino_full_theta} but for $E=0.6$~GeV.
}
}
\end{figure*}
\begin{figure*}
\centering{
	\includegraphics[width=0.9\textwidth]{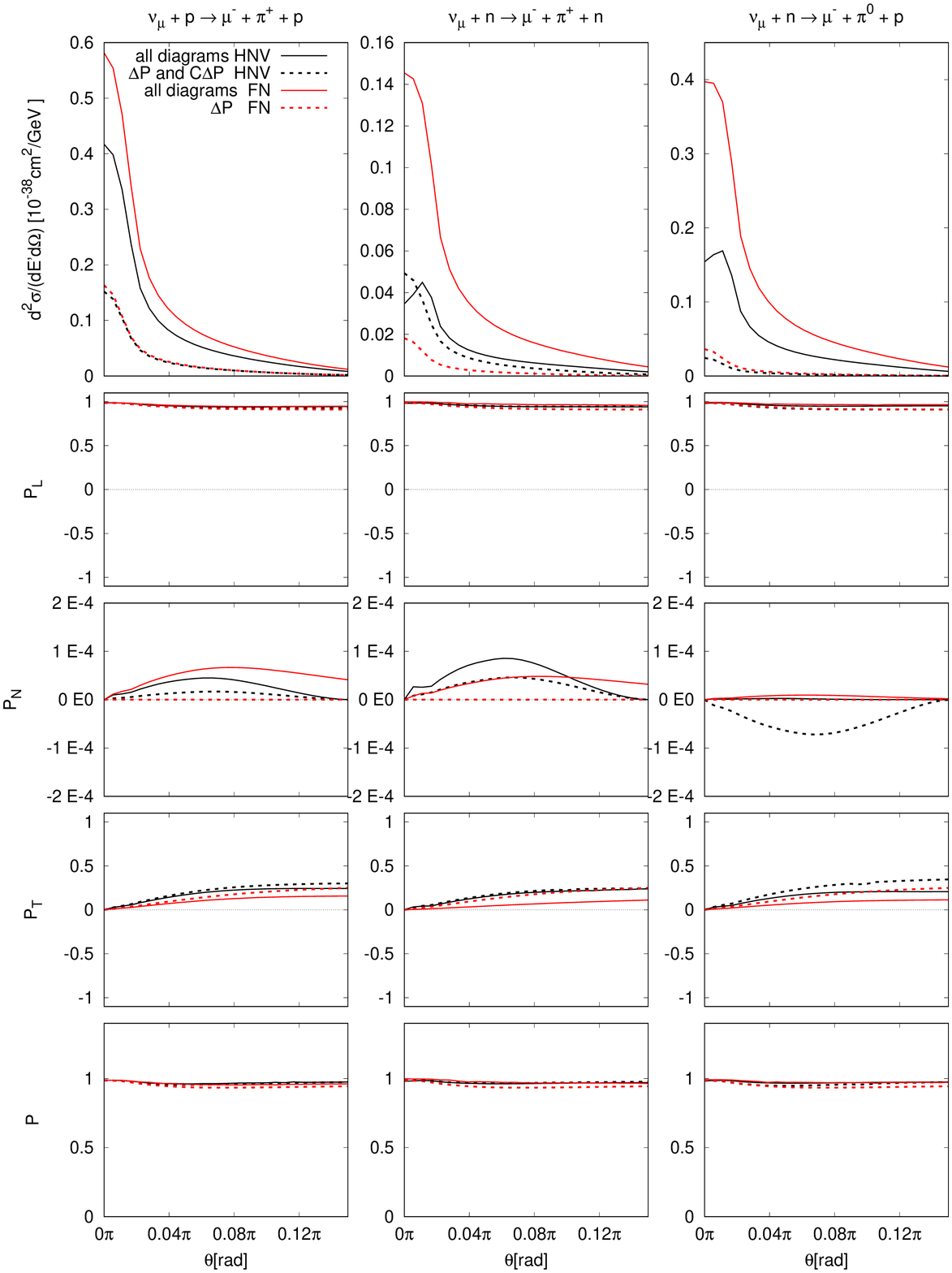}
\caption{\label{Fig_muon_neutrino_full_theta_t2k} 
Caption the same as in Fig. \ref{Fig_muon_neutrino_full_theta} but the predictions are T2K-flux averaged.
}
}
\end{figure*}


\begin{figure*}
\centering{
\includegraphics[width=\textwidth]{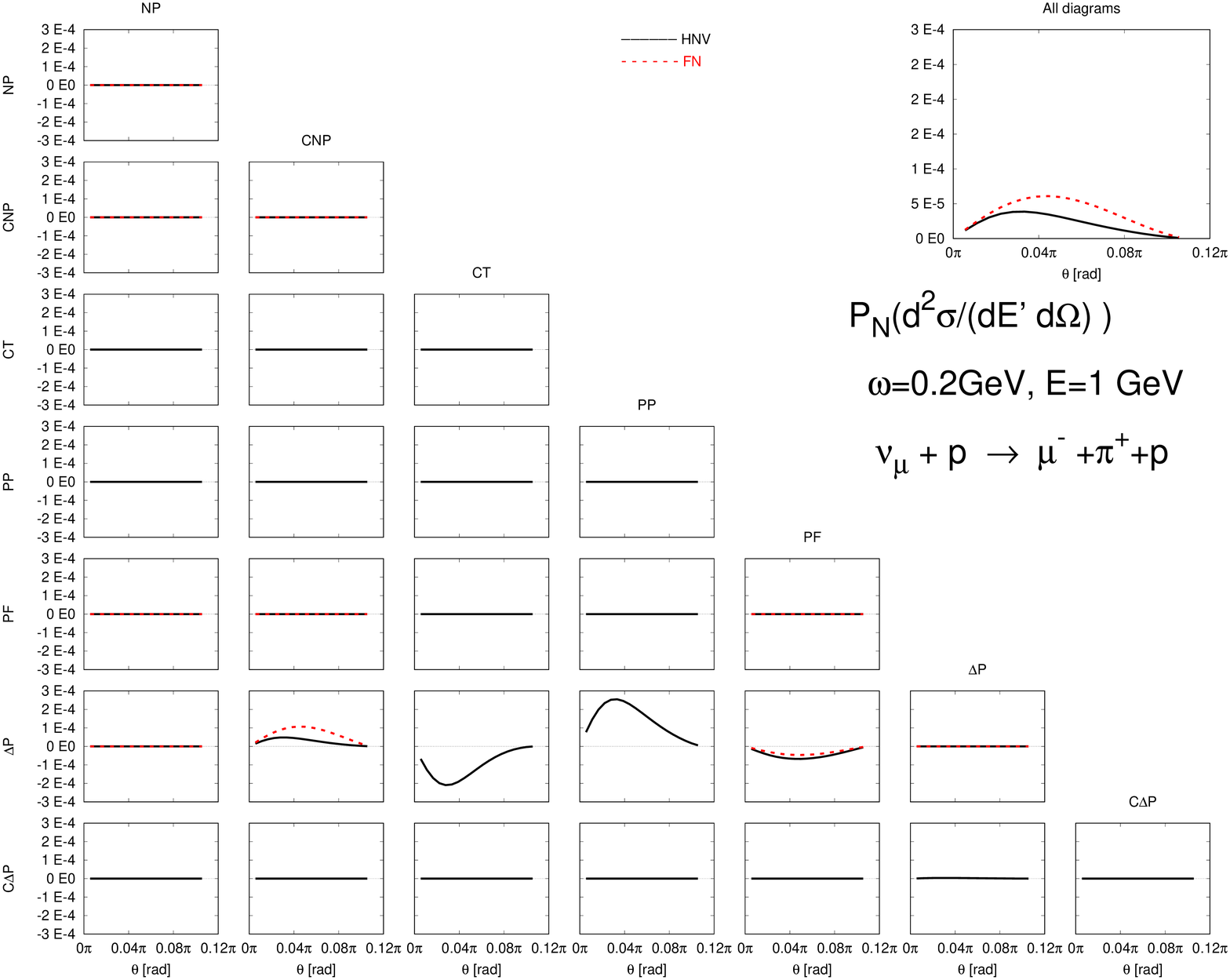}
\caption{\label{Fig_muon_neutrino_normal_component_partition}
Separation of the normal component of the polarization of $\mu^-$ into various interference contributions.  The dotted/solid (black/red) line represents the HNV/FN model predictions.  The contribution from the $ |\mathcal{M}_a|^2 $  are on the diagonal while below the diagonal the interference terms  $2\Re (\mathcal{M}_i \mathcal{M}_j^*)$ ($j$ -- column, $i$ -- row) are plotted. 
}
}
\end{figure*}

\section{SPP Models}
\label{Section_Models}

We consider the interactions of the neutrinos of energy of the order of 1~GeV  with the free nucleon target. The dominant RES contribution is given by the weak $N \to \Delta(1232)$ transition. 
 In both the HNV and the FN models the $N \to \Delta (1232)$ transition is described by the same formalism but in the FN model the weak vertex $N W^+ \Delta$ is oversimplified. In  phenomenological approaches the NB contribution is described by some number of diagrams allowed by a symmetry, however, in the FN model the number of diagrams is smaller.   In next two subsections the main features of both  descriptions are shortly reviewed.

\subsection{HNV Model}

We consider the HNV model as it is formulated in~\cite{Hernandez:2007qq}, however, we notice the latest developments of the approach in Refs. \cite{Alvarez-Ruso:2015eva,Hernandez:2016yfb}. The NB contribution is described by five diagrams, see Fig. \ref{Fig_diagrams}. They  are determined from the non-linear sigma model.   The main idea was to obtain, on the tree level, the axial and the vector currents from the non-linear sigma model and to associate them with their phenomenological counterparts. The resonance contribution is given by two diagrams describing the weak $N\to \Delta(1232)$ transition.  The  $\Delta(1232)$ resonance is modeled by the $3/2$-spin Rarita-Schwinger field~\cite{Rarita:1941mf}.

\subsubsection{Nonresonant Background}

The NB contribution is generated by five hadron currents: 
\begin{eqnarray}
\label{Current_IFIC_NP}
r^{\mu}_{NP}
&=&
-
 \frac{i g_A}{\sqrt{2} f_{\pi}}
\slashed{k}_{\pi} \gamma_5 
S_N(p+q)
\Gamma_N^\mu (q) 
\\
\label{Current_IFIC_CNP}
r^{\mu}_{CNP} & = &
-
  \frac{i g_A}{\sqrt{2} f_{\pi}}
  \Gamma_N^\mu (q)
S_N(p'-q)
\slashed{k}_{\pi} \gamma_5   
\\
\label{Current_IFIC_CT}
r^{\mu}_{CT} &=&
-
  \frac{i}{\sqrt{2} f_{\pi}}
 \gamma^{\mu} (g_A  F_1^V (q) \gamma_{5} - F_{\rho}(q-k_{\pi})) 
\\
\label{Current_IFIC_PP}
r^{\mu}_{PP} &=&
     -\frac{i }{\sqrt{2} f_{\pi}}
F_{\rho}(q-k_{\pi})
\frac{q^{\mu}}{q^2 - m_{\pi}^2} \slashed{q} 
\\
\label{Current_IFIC_PF}
r^{\mu}_{PF} &=&
-    F_1^V \frac{2 i g_A M}{\sqrt{2} f_{\pi}}
\frac{2k_{\pi}^{\mu}-q^{\mu}}{(k_{\pi}-q)^2 - m_{\pi}^2}  \gamma_5 
\end{eqnarray}
where 
\begin{equation}
    S_N(P) = \frac{\slashed{P}+M}{P^2-M^2}
\end{equation}
is the nucleon propagator. 

The electroweak nucleon vertex reads
\begin{eqnarray}
\Gamma_N^{\mu}(q) 
&=& 
F_1^V(q) \gamma^{\mu}+i F_2^V(q) \frac{\sigma^{\mu \nu} q_{\nu}}{2M}
\nonumber \\
& & - 
G_A(q) \left(\gamma^{\mu}+  \frac{\slashed{q} q^{\nu}}{m_{\pi}^2-q^2} \right)\gamma_5,
\end{eqnarray} 
where $F_{k}^V(q^2)$ is the vector nucleon form factor, see (\ref{vector_form_factor}); $G_A$ is the axial nucleon form factor (see: \ref{axial_form_factor}); $g_A=1.26$ is the axial nucleon coupling; the form factor $F_\rho$  is given by (\ref{Frho}).  

\subsubsection{$\Delta(1232)$ Contribution}

The $N \to \Delta(1232)$ resonance transition  currents have the form:
\begin{eqnarray}
\label{Current_IFIC_DP}
r^{\mu}_{\Delta P} &=& i 
 \sqrt{3} \, \frac{f^*}{m_\pi}
   k^\alpha_\pi S_{\alpha\beta}(p+q)
\Gamma ^{\beta\mu}(p,q) 
\\
\label{Current_IFIC_CDP}
r^{\mu} _{C\Delta P} 
&=& \!\!\!
 \frac{i}{\sqrt{3}} \frac{f^*}{m_\pi}   \gamma^0 [\Gamma ^{\alpha\mu }(p',-q)]^\dagger \gamma^0  k^\beta_\pi S_{\alpha\beta}(p'-q) 
\end{eqnarray}
where $f^* = 2.14$.
\begin{eqnarray}
\nonumber
 S_{\alpha\beta}(p)&=&-\frac{(\slashed{p}+M_{\Delta})}{p^2 - M_\Delta^2 + i M_\Delta \Gamma_{\Delta}(p) }
 \\
 & & \!\!\!\!\!\times \nonumber
 \left( g_{\alpha\beta} -\frac{1}{3}\gamma_{\alpha}\gamma_{\beta}- \frac{2}{3}\frac{p_{\alpha}p_{\beta} }{M_{\Delta}^2} + \frac{1}{3}\frac{\displaystyle p_{\alpha}\gamma_{\beta}-\gamma_{\alpha}p_{\beta} }{M_{\Delta}}
\right),
\\
\label{3_2_spin_propagator}
\end{eqnarray}
is $3/2$-spin particle propagator; $M_\Delta = 1232$~MeV, while $\Gamma_{\Delta}(p)$ is the resonance width, which in the HNV model is given by (\ref{Width_HNV}).

The electroweak vertex for $W^+ (q) N(p) \to \Delta (p+q) $ transition has the vector$+$axial form, namely
\begin{equation}
\label{Gamma_V_A}
{\Gamma^{\nu\mu}}(p,q)=\left[V^{\nu\mu}_{3/2}(p,q) + A^{\nu\mu}_{3/2}(p,q) \right]\gamma_5.
\end{equation}   
The vector part of (\ref{Gamma_V_A}) reads \cite{Jones:1972ky} 
\begin{eqnarray}
V^{\nu\mu}_{3/2}(p,q)&=&
\frac{C_3^V}{M}(g^{\nu\mu} \slashed{q} - q^{\nu}\gamma^{\mu})
\nonumber \\
& + &
\frac{C_4^V}{M^2}(g^{\nu\mu} q\cdot (p+q) - q^{\nu}({p}^{\mu}+q^\mu))
\nonumber \\
& +& 
\frac{C_5^V}{M^2}(g^{\nu\mu} q \cdot p - q^{\nu}p'^{\mu}).
\end{eqnarray}
The vector form factors are given by the fit form \cite{Graczyk:2014dpa}, see Appendix~\ref{Appemdix_ND_TransitionFormFactors}.

The axial part of the vertex (\ref{Gamma_V_A}) reads~\cite{Schreiner:1973mj} 
\begin{eqnarray}
A^{\nu\mu}_{3/2}(p,q)&=&
\left( \frac{C_3^A}{M}(g^{\nu\mu} \slashed{q} - q^{\nu}\gamma^{\mu})
\right. \nonumber \\
& & 
\left. +
\frac{C_4^A}{M^2}(g^{\nu\mu} q\cdot (p+q) - q^{\nu}({p}^{\mu}+{q}^{\mu}))
\right. \nonumber \\
& &\left. +g^{\nu\mu}C_5^A
+ \frac{C_6^A}{M^2}q^{\nu}q^{\mu} 
\right)\gamma_5.
\end{eqnarray}
The axial form factors are obtained from the  analysis of the neutrino scattering  data. However, the data are not enough accurate to get an information about four independent form factors~\cite{Graczyk:2009zh}. 
Therefore the following simplifications are made: 
\begin{itemize} 
\item  $ C_3^A(Q^2)=0$, as it is suggested by simple quark model, see e.g. \cite{Graczyk:2007bc}; 
\item  
\begin{equation}
C_4^A(Q^2) = - \frac{C_5^A(Q^2)}{4},
\label{C4_C5}
\end{equation}
as it is supported by dispersion analyses \cite{Adler:1968tw}; 
\item  $C_6^A$ is a function   of $C_5^A$, namely: 
\begin{equation}
C_6^A(Q^2) =   \frac{M^2}{m_\pi^2 + Q^2} C_5^A(Q^2).
\end{equation}
as it is obtained from  the partially conserved vector current (PCAC) hypothesis.
\end{itemize}
For the numerical analyses we take the $C_5^A$ fit from \cite{Graczyk:2014dpa}, see Appendix~\ref{Appemdix_ND_TransitionFormFactors}.


\subsection{ Fogli-Nardulli Model}

The FN model\footnote{Short review of the FN model can be found in \cite{Zmuda_thesis}.} formulated  in \cite{Fogli:1979cz} describes the pion production in the first and the second  resonance regions. But in  this paper  we consider only the first resonance region.

\subsubsection{Nonresonant Background }

The NB contribution is described by three diagrams: the  pion in flight ($PF$), the same as in the HNV model, and  two  nucleon pole diagrams: $NP$, and $CNP$. But in the latter the pseudoscalar pion-nucleon coupling was implemented, while in the HNV model the pseudovector pion-nucleon coupling is discussed.
\begin{eqnarray}
\tilde r_{NP}^{\mu} &=& i\sqrt{2} g_{NN\pi} \gamma_5 S_N(p+q)  \Gamma^\mu_N(q)
\\
\tilde r_{CNP}^{\mu} &=& i\sqrt{2} g_{NN\pi}   \Gamma_N^\mu(q) S_N(p'-q)  \gamma_5 
\\
\tilde r_{PF}^{\mu} &=& -i\sqrt{2} g_{NN\pi}  \gamma_5 \frac{2k_{\pi}^\mu-q^\mu }{(k_{\pi}-q)^2-m_{\pi}^2} F_{\pi}(k_\pi - q),
\end{eqnarray} 
where $g_{NN\pi} = \sqrt{14.8 \cdot 4 \pi}$, see e.g. \cite{deSwart:1997ep}, $F_\pi$ form factor is given in the Appendix \ref{Appemdix_ND_TransitionFormFactors}. 
\begin{figure}
   \centering{
\includegraphics[width=0.37\textwidth]{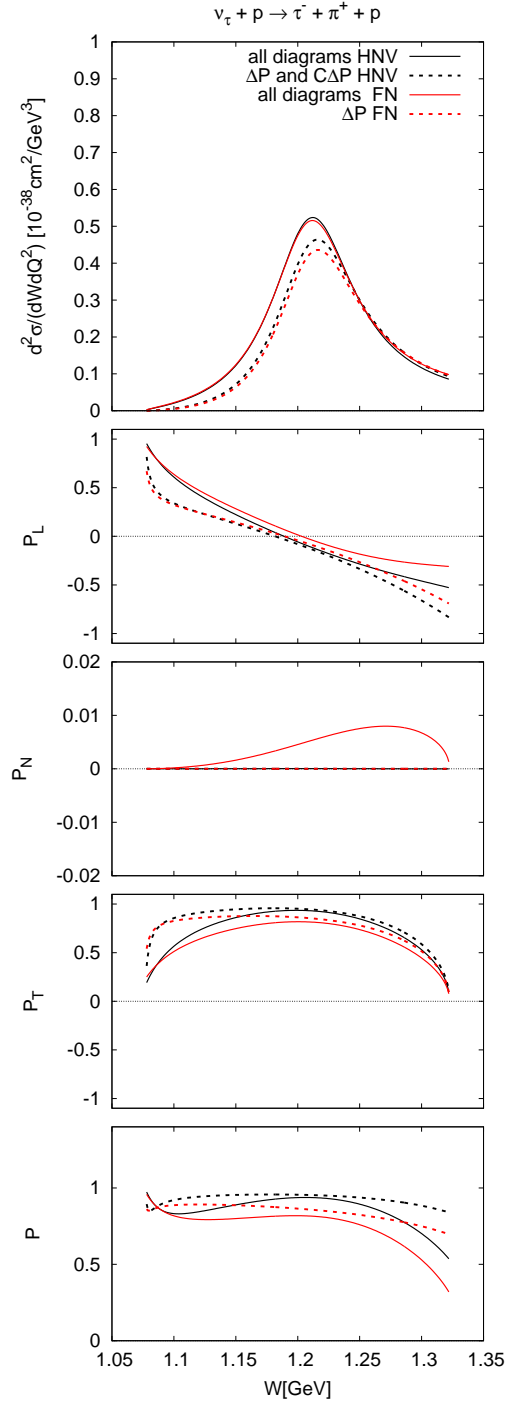}
	\caption{
	$W$-dependence of the components of  $\mathcal{P}_{\tau^-}^\alpha(E,Q^2,W)$ vector polarization and the degree of polarization (last row) of the $\tau^-$ lepton, for the energy of the neutrino $E=5$~GeV and   $Q^2=1$~GeV$^2$.
		The dotted/solid line represents the RES/full model contributions of the HNV (black) and   the FN (red) models. The resonance contribution is given by $|\mathcal{M}_{\Delta P} + \mathcal{M}_{C\Delta P}|^2$ and $|\mathcal{M}_{\Delta P}|^2$ in the HNV and the FN models respectively.
	\label{Fig_tau_dwdQ2} }
	}
\end{figure}

\subsubsection{$\Delta(1232)$ Contribution}

\begin{figure*}
\centering{
\includegraphics[width=0.9\textwidth]{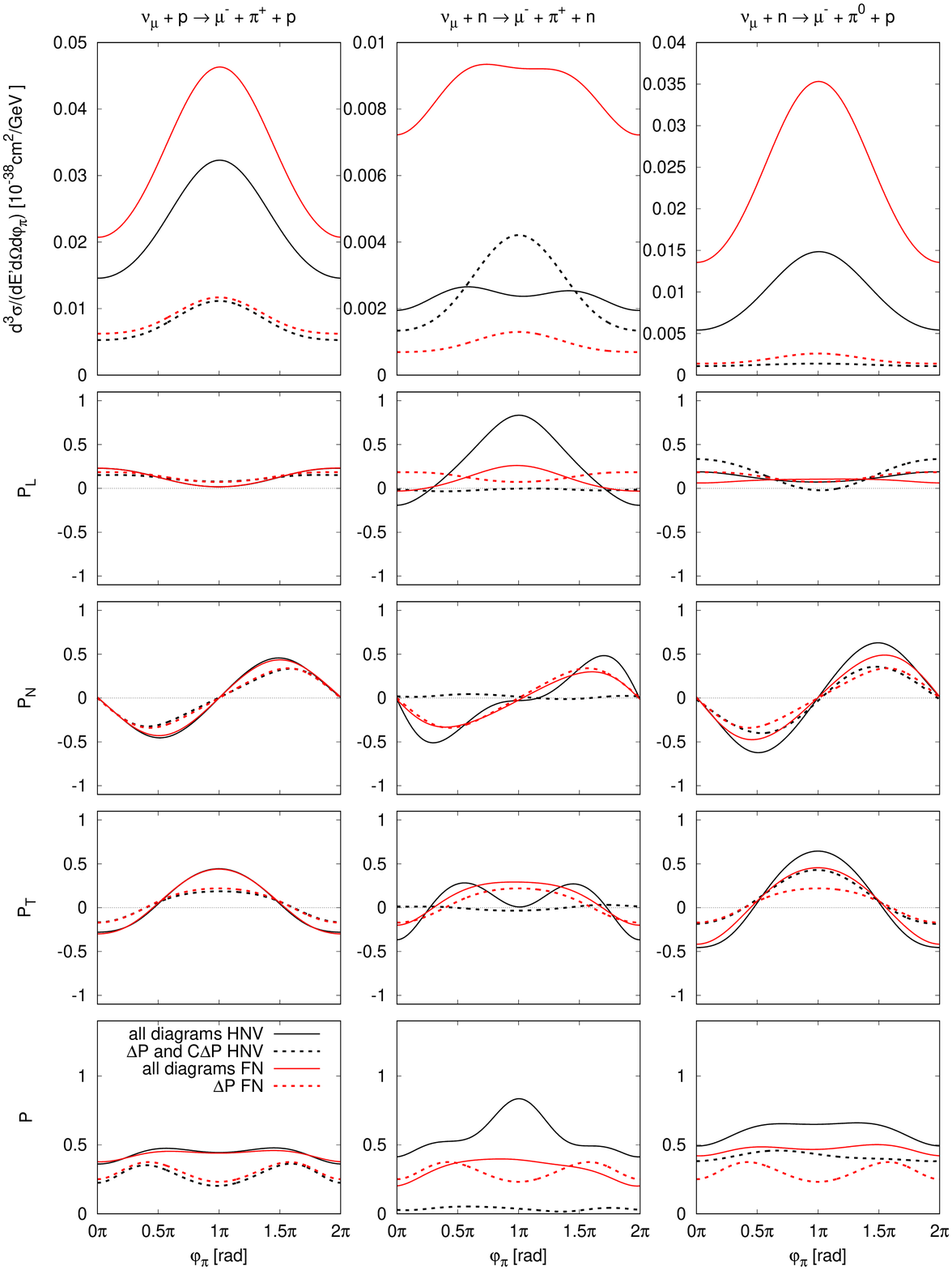}
	\caption{\label{Fig_Polaryzacje_neutrina_Nukleon_omega=05theta=5} 
Angular dependence of the components of  $\mathcal{P}_{N}^\alpha(E,\theta,\omega, \phi_\pi)$ vector polarization and the degree of polarization (last row) of the nucleon, for the energy of the neutrino $E=1$~GeV, the scattering angle $\theta=5^{\circ}$ and the energy transfer $\omega=0.2$~GeV.
		The dotted/solid  line represents the RES/total contributions of the HNV (black) and   the FN (red) models. The resonance contribution is given by $|\mathcal{M}_{\Delta P} + \mathcal{M}_{C\Delta P}|^2$ and $|\mathcal{M}_{\Delta P}|^2$ in the HNV and the FN models respectively.
		}
	}
\end{figure*}
\begin{figure*}
\centering{
\includegraphics[width=0.9\textwidth]{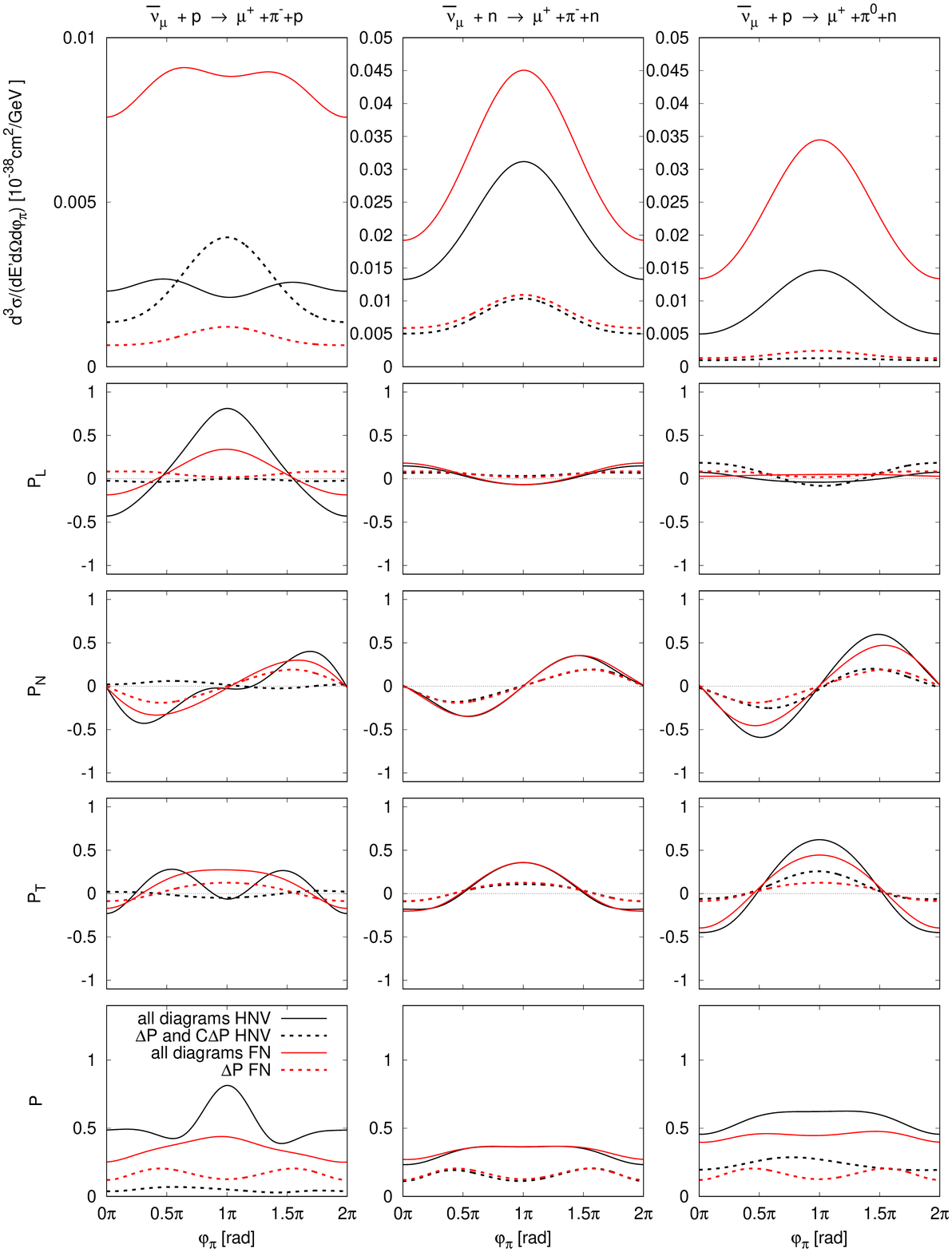}
\caption{\label{Fig_Polaryzacje_antyneutrina_Nukleon_omega=05theta=5} 
	The same caption  as in Fig.~\ref{Fig_Polaryzacje_neutrina_Nukleon_omega=05theta=5} but for $\overline \nu_\mu N$ scattering. 
		}
}
\end{figure*}
\begin{figure*}
\centering{
\includegraphics[width=0.9\textwidth]{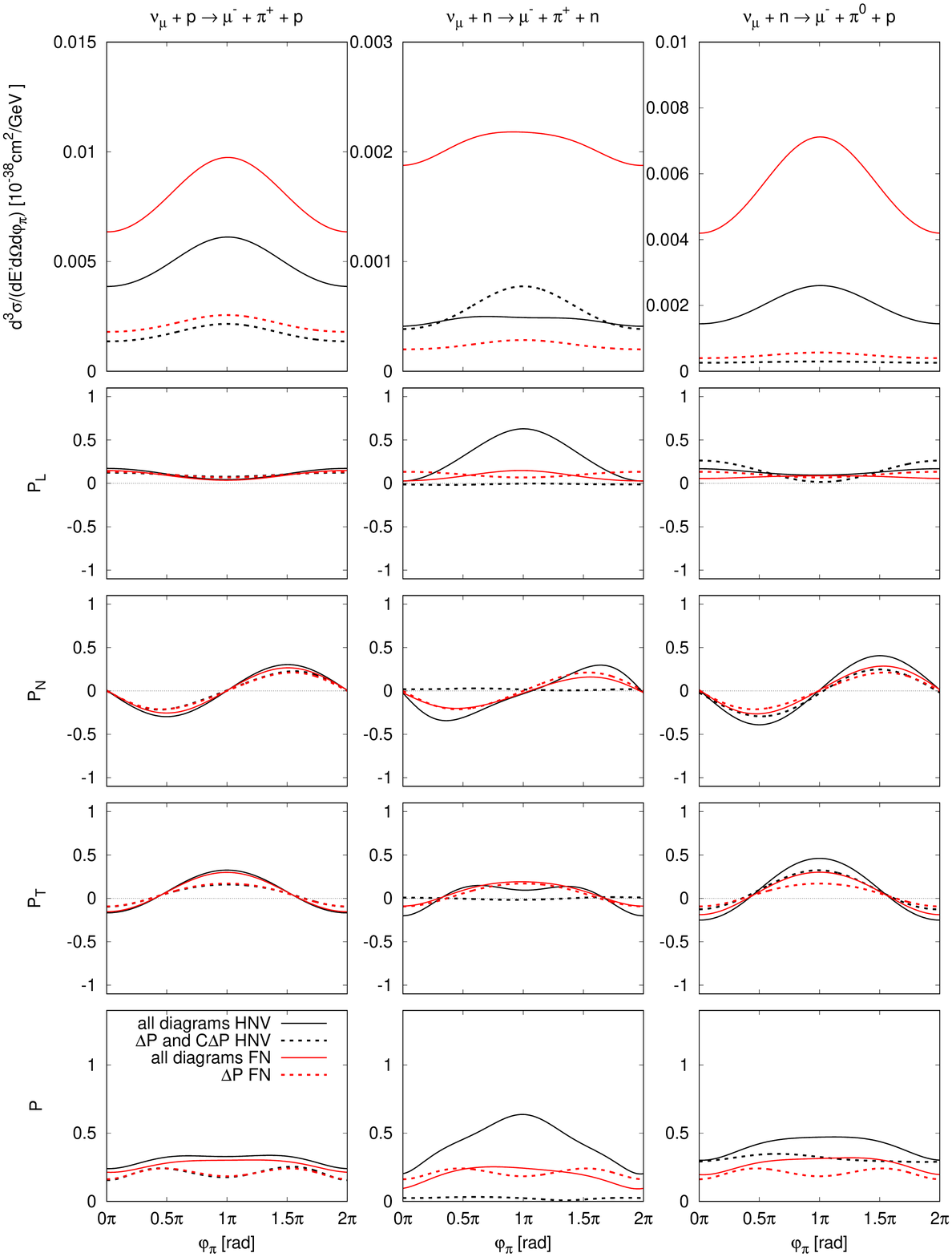}
	\caption{\label{Fig_Polaryzacje_neutrina+E=06_Nukleon_omega=05theta=5} 
	The same caption  as in Fig.~\ref{Fig_Polaryzacje_neutrina_Nukleon_omega=05theta=5} but for the neutrino energy $E=0.6$~GeV and the energy transfer $\omega=0.2$~GeV.
		}
	}
\end{figure*}
\begin{figure*}
\centering{
\includegraphics[width=0.9\textwidth]{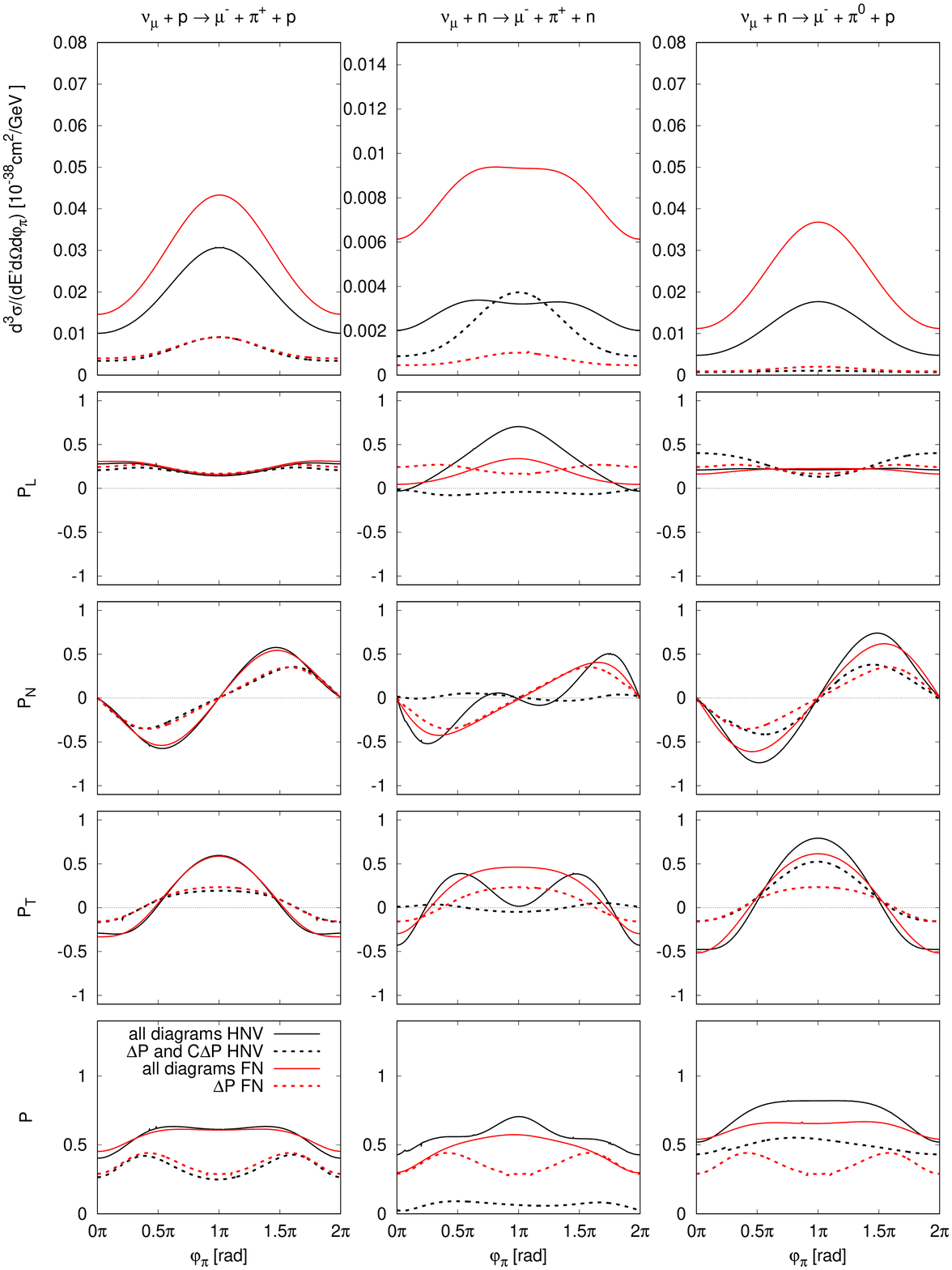}	
\caption{
\label{Fig_Polaryzacje_neutrina_Nukleon_omega=05theta=5_t2kflux} 
	The same caption  as in Fig.~\ref{Fig_Polaryzacje_neutrina_Nukleon_omega=05theta=5} but T2K-flux averaged. 
		}
}
\end{figure*}

In the FN model only one digram describes the SPP induced by  the $N \to \Delta(1232)$ transition. The structure of this current is the same as  in the HNV model 
\begin{eqnarray}
	\tilde{r}_{\Delta N}^{\mu} & \to & r_{\Delta N}^{\mu}.
\end{eqnarray} 
However, it is assumed that 
\begin{equation}
C_5^V(Q^2)=C_4^A(Q^2) = C_3^A(Q^2) = 0
\end{equation} 
as well as 
\begin{equation} 
C_4^V(Q^2) = - \frac{M}{W} C_3^V.
\end{equation}
The width of the $\Delta(1232)$ resonance is given by (\ref{Width_FN}).

\section{Numerical Results}
\label{Section_Results}

\subsection{Unpolarized Cross Sections}

\begin{figure*}
	\centering{
\includegraphics[width=0.9\textwidth]{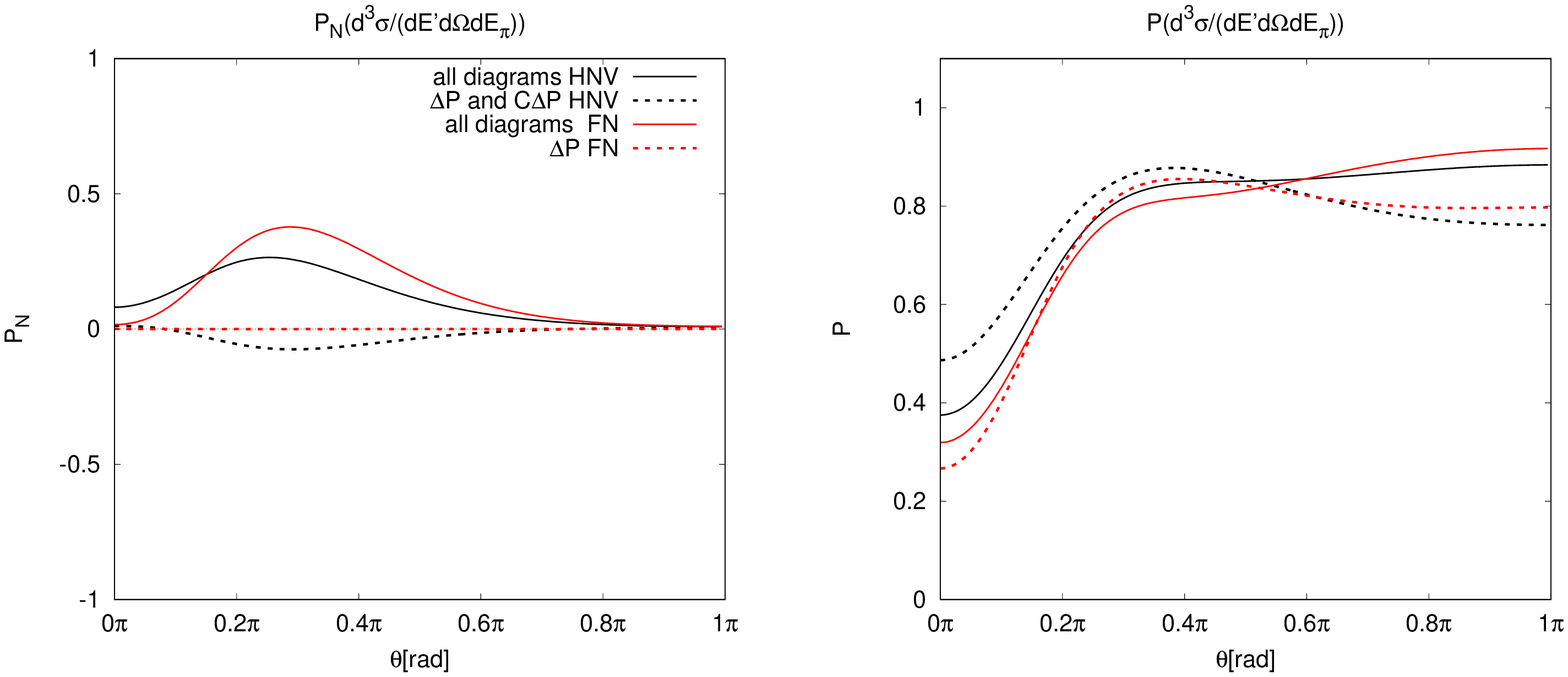}
		\caption{
			\label{Figure_Normal_Component_of_Nucleon_Partition}
			The angular dependence  in $\theta$  of the normal component of the polarization (left figure) and the degree of polarization (right figure) of the recoiled proton in the process $\nu_\mu + p \to \mu^- + p + \pi^+$. The predictions are obtained within HNV  (black) and FN (red) models. The RES/full model contribution is denoted by dotted/solid line. The neutrino energy $E=0.7$~GeV, the energy transfer $\omega=0.5$~GeV,  and the pion energy $E_\pi=0.25$~GeV. The resonance contribution is given by $|\mathcal{M}_{\Delta P} + \mathcal{M}_{C\Delta P}|^2$ and $|\mathcal{M}_{\Delta P}|^2$ in the HNV and the FN models respectively.
		} 
	}
\end{figure*}
Most of the SPP models, including the HNV and the FN approaches, reproduce  the  cross section data for  $\nu_\mu p \to \mu^- p \pi^+$ process with reasonable accuracy. The agreement with the data is achieved by appropriate tuning the values of  the parameters of the $C_5^A$ axial form factor \cite{Fogli:1979cz,Hernandez:2010bx,Lalakulich:2010ss,Graczyk:2014dpa,Zmuda_thesis}. 
The main problem is to obtain  coherent model predictions for all three CC  channels. Indeed the ANL data for  
the  $\nu n \to \mu^- n \pi^+$ process seem to be inconsistent with the other channels~\cite{Graczyk:2014dpa}. This can be  caused by:   oversimplified treatment of the deuteron structure effects in the analysis of the ANL data (see \cite{Wu:2014rga})    and/or incomplete description of the $\Delta(1232)$ resonance propagation (see \cite{Hernandez:2016yfb}) and/or  low quality of the data.

It seems that critical studies of the SPP models can be performed only if  new more precise measurements of the interactions of the neutrinos with the free nucleon target will be delivered. Moreover the information about the SPP dynamics hidden in: the  total and single differential cross sections (like $d\sigma /dQ^2$) and even double differential cross sections is limited because important features of an approach are integrated out. Hence the analysis of these cross section  data does not allow to verify which model, among many in the market, is the closest, in the description, to  the reality.  It is illustrated in Fig. \ref{Fig_partition_of_the_nonresonant}, where  we plot  $d^2\sigma/dWdQ^2$ distribution and its partition into: 
\begin{enumerate}[(i)]
\item ''pure'' resonance contribution;

\item interference between the RES and the NB amplitudes;

\item ''pure'' NB contribution.
\end{enumerate}      
Although the RES and the NB contribute differently in both models the shape and the magnitude of $d^2\sigma/dW dQ^2$  obtained within both approaches are very similar.  The significant disparities between predictions of the both models manifest when the triple differential cross sections are examined. In Fig.~\ref{Fig_cross_section_partition} we plot $d^3\sigma/dE'd\Omega d E_\pi$ and its partition into all possible interference terms calculated at particular kinematics. At low scattering angle the FN cross section increases rapidly while the HNV model predictions fall down.  The low-$\theta$ behavior of the FN's cross section is determined by a $|\mathcal{M}_{NP}|^2$ contribution.  In the HNV model the contribution from the $NP$ diagram  is smaller and it is suppressed by the contribution missing in the FN model, namely the interference terms: $\mathcal{M}_{C\Delta P} \mathcal{M}_{NP}^*$, $\mathcal{M}_{CT} \mathcal{M}_{PF}^*$ and others.  

Certainly inspection of the triple differential cross sections, their dependence on $\phi_{\pi}$ angle, may deliver a valuable  information about the RES and the NB dynamics. Let us also mention that  additional constraints  on the NB contribution can be obtained  from the combined analysis of various mass  distributions. It is demonstrated in  Ref. \cite{Lalakulich:2010ss}, where the detailed discussion of $W(N \pi)$, $W(\mu N)$ and $W(\mu \pi)$ event distributions of the ANL and the BNL experiments were performed.   However, in the next section we shall demonstrate that the  PT observables contain an unique information about the dynamics of the SPP.

\subsection{Polarization Transfer Observables}


We start the presentation of the PT results from the discussion of the polarization properties of the charged lepton produced in  process (\ref{Process_lepton_polarization}). 
In Figs. \ref{Fig_muon_neutrino_full_theta} and \ref{Fig_muon_antyneutrino_full_theta} the longitudinal, normal and transverse components as well as the degree of polarization of the $\mu^-$ and $\mu^+$ leptons are shown respectively. Additionally  Fig.~\ref{Fig_muon_neutrino_E=06_full_theta} includes the plots of $\mu^-$ polarization components calculated for an averaged T2K \cite{Abe:2011ks} experiment neutrino energy, $E=0.6$. While in Fig.~\ref{Fig_muon_neutrino_full_theta_t2k} the polarization components of $\mu^-$ lepton  averaged over the T2K energy flux \cite{Abe:2012av} are shown. The muon lepton is a light particle hence  it should be almost polarized. However, we notice that for some kinematics, namely  at low scattering angle, the $\mu^\pm$ lepton can be partially polarized.

A  unique information about the SPP dynamics is hidden in the normal component of the polarization. Indeed this observable  is dominated by the interference of the $\mathcal{M}_{\Delta P}$ and $\mathcal{M}_{C\Delta P}$ amplitudes with the NB diagrams. Therefore the sign of the normal component is defined by  the   relative sign between the RES and the NB contributions. This property is illustrated in Fig. \ref{Fig_muon_neutrino_normal_component_partition}, where we plot the partition of the normal component into all interference terms.

Now let as review the polarization properties  the $\tau$ lepton produced in the $\nu_{\tau}N $ scattering. The mass of the $\tau$ is large  hence it can  be partially polarized. Indeed in Fig.~\ref{Fig_tau_dwdQ2} we plot the predictions of the degree of polarization. It varies from $0.3$ to $1.0$. Let us remark that the polarization vector of the $\tau$ lepton   is one of the variables which describes the angular distribution of the products of its decay. The model dependence of 
the predictions of the polarization vector of the tau lepton on the SPP model assumptions seems to be  stronger  then in the case of muon lepton.

Interesting conclusions are obtained from the analysis of the polarization properties of the final nucleon produced in process (\ref{Process_final_nucleon_polarization}). In Figs. \ref{Fig_Polaryzacje_neutrina_Nukleon_omega=05theta=5} and \ref{Fig_Polaryzacje_antyneutrina_Nukleon_omega=05theta=5} the angular dependence (in $\phi_{\pi}$ angle) of the polarization components and the degree of polarization of the nucleon produced in $\nu_\mu N$ and $\overline{\nu}_\mu N$ interactions are plotted. Similarly as for the discussion of the lepton polarization properties we show also the predictions of the polarization components: for the T2K neutrino averaged  energy, Fig.~\ref{Fig_Polaryzacje_neutrina+E=06_Nukleon_omega=05theta=5}, and   T2K-flux averaged, Fig.~\ref{Fig_Polaryzacje_neutrina_Nukleon_omega=05theta=5_t2kflux}.
Notice that the resonance part of the normal, transverse and longitudinal polarization components have sinusoidal character which is distorted by the inclusion of the NB contribution. 
At some kinematics the normal polarization of the nucleon is large -- it reaches the value $0.5$. Similarly as in the lepton polarization case the interference of the RES with the NB diagrams gives sizeable contribution which for some kinematics becomes dominant. It is illustrated in Fig.~\ref{Figure_Normal_Component_of_Nucleon_Partition}.

\section{Summary}
\label{Section_Summary}

We discussed  polarization properties of fermions produced in the SPP processes induced by the charged current neutrino-nucleon scattering. The components of the polarization vector  of the outgoing charged lepton and the nucleon  were calculated.  In order to make the discussion more realistic we have made the predictions of the polarization properties for the T2K experiment.    It turned out that the PT observables are very sensitive on the details of the SPP model. 
   In order to investigate how a change in a model assumptions affects the predictions of the polarization components  we considered two the SPP approaches.  It was demonstrated that the most interesting information about the SPP dynamics is hidden in the normal component of the polarization of the outgoing lepton and the nucleon. In particular the sign of the normal polarization of the charged lepton is determined by the relative sign between the NB and the RES amplitudes. 
   
   Eventually we conclude that  the investigation of the polarization transfer observables in the SPP in $\nu N$ scattering should deliver a complementary,  to the spin averaged cross sections,  knowledge about the resonance and the nonresonance contributions.

\begin{acknowledgments}
All  algebraic calculations have been performed using FORM symbolic language \cite{Vermaseren:2000nd}.

The calculations have been carried out in Wroclaw Centre for
Networking and Supercomputing (\url{http://www.wcss.wroc.pl}),
grant No. 268.
\end{acknowledgments}

\appendix

\section{Normalization}

The Dirac field of $1/2$-particle of mass $M$ and momentum $p$ is normalized so that:
\begin{eqnarray}
\overline{u}(p,s) u(p,s') &=& 2 M \delta_{ss'}
\\
u(p,s) \overline{u}(p,s) &=& \frac{1}{2}\gamma_5 \slashed{s}(\slashed{p}+M),
\end{eqnarray}
where $\slashed{p} \equiv p_\mu \gamma^\mu$. 

One particle fermion/scalar state is normalized so that
\begin{equation}
\langle \mathbf{p'},s'\mid  \mathbf{p},s\rangle = 2 E_p (2\pi)^3\delta_{ss'}\delta^{(3)}(\mathbf{p}-\mathbf{p'}),
\end{equation}
where $\mid \mathbf{p},s\rangle \equiv a^\dagger(\mathbf{p},s)\mid 0 \rangle$, $a^\dagger$ is the creation operator of particle with momentum $\mathbf{p}$ and spin $s$.

	 		 					\begin{table}[h]
	 		 						\begin{center}
	 		 							\begin{tabular}{|c|c|c|c|c|c|c|c|}
	 		 								\hline
	 		 								process & NP & CNP & CT & PP& PF & $\Delta P$ & $C\Delta P$\\
	 		 								\hline
	 		 								& & & & & & &\\
	 		 								
	 		 								$\nu_l  p \to  l^-   p  \pi^+$        &  0  &  $1$   &  $1$  & $1$   & $1$    &  $ 1 $    &  $1 $    \\
	 		 								& & & & & & &\\
	 		 								
	 		 								$\nu_l  n \to  l^-   p  \pi^0$        &  $\displaystyle \frac{1}{\sqrt{2}}$  &  $\displaystyle -\frac{1}{\sqrt{2}}$    & $-\sqrt{2}$  &  $-\sqrt{2}$  &    $-\sqrt{2}$  &         $ -\displaystyle \frac{\sqrt{2}}{3}$        &   $ \sqrt{2 } $            \\
	 		 								& & & & & & &\\

	 		 								$\nu_l  n \to  l^-   n  \pi^+$        &  $1$  &  0   &   $-1$  &  $-1$   &  $-1$   &  $\displaystyle \frac{1}{3}$    &  $  3$                   \\
	 		 								& & & & & & &\\
	 		 								$\overline{\nu}_l  n \to  l^+   n  \pi^-$      &  0  &  $1$   &  $1$  & $1$   & $1$    &  $ 1 $    &  $1 $    \\
	 		 								& & & & & & &\\
	 		 								$\overline{\nu}_l  p \to  l^+   n  \pi^0$         &   $\displaystyle -\frac{1}{\sqrt{2}}$  &   $\displaystyle\frac{1}{\sqrt{2}}$    &   $\sqrt{2}$  &  $\sqrt{2}$  &    $\sqrt{2}$  &         $ \displaystyle \frac{\sqrt{2}}{3}$       &   $- \sqrt{2 } $             \\
	 		 								& & & & & & &\\
	 		 								$\overline{\nu}_l  p \to  l^+   p  \pi^-$       &  $1$  &  0   &   $-1$  &  $-1$   &  $-1$   &  $\displaystyle\frac{1}{3}$   &  $  3$                   \\
	 		 								& & & & & & &\\
	 		 								\hline
	 		 							\end{tabular}
	 		 						\end{center}
	 		 						\label{Clebsch-Gordan-CC}
	 		 						\caption{ Clebsch-Gordan coefficients.}
	 		 					\end{table}

\section{Transition Form Factors}

The vector nucleon form factors $F_1^V$ and $F_2^V$ are expressed in terms of electromagnetic neutron, $F_k^n$, and proton, $F_k^p$, form factors, namely 
\begin{equation}
\label{vector_form_factor}
F_{k}^V(q) = F_k^p(q) - F_k^n(q), \; k=1,2.
\end{equation}
We consider the same nucleon form factors  as in \cite{Hernandez:2007qq}.  

The axial nucleon form factor reads
\begin{equation}
\label{axial_form_factor}
G_A(q)  = \frac{g_A}{\displaystyle \left(1 - \frac{q^2}{M_A^2} \right)^2}, 
\end{equation}
where $M_A = 1.00$ GeV, $g_A  = 1.26$.

\subsection{HNV Model}
\label{Appemdix_ND_TransitionFormFactors}

The vector $N\to \Delta(1232)$ transition form factors are given by~\cite{Graczyk:2014dpa}:
\begin{eqnarray}
\label{eq:vector_Delta_Form_Factors}
C_3^V(Q^2)&=&\frac{C_3^V(0)}{1+AQ^2+BQ^4+CQ^6}\cdot(1+K_1Q^2),\nonumber 
\\\\
C_4^V(Q^2)&=&-\frac{M}{W}C_3^V(Q^2)\cdot\frac{1+K_2Q^2}{1+K_1Q^2},\\
\label{eq:c5K} C_5^V(Q^2)&=&\frac{C_5^V(0)}{\displaystyle \left(1+D\frac{Q^2}{M_V^2}\right)^2},
\end{eqnarray} 
where $M_V=0.84$~GeV, parameters  $K_1$, $K_2$, $A$, $B$, $C$ and $D$ are in Table \ref{Table_Cparameters}.
\begin{table}[h]
	\begin{tabular}{|cccccccc|}
		\hline
		$C_3^V(0)$ & $C_5^V(0)$ & $A$ & $B$ & $C$ & $D$ & $K_1$ & $K_2$\\
		\hline
		$\, 2.10$
		& $\, 0.63$
		& $\, 4.73$
		& $-0.39$
		& $\,5.59$
		& $\, 1.00$
		& $\, 0.13$
		& $\, 1.68$\\
		\hline
	\end{tabular}
	\label{tab:ourfitt}
	\caption{\label{Table_Cparameters} Parameters of the vector form factors for the $N \to \Delta$ transition.}
\end{table}

The axial form factor $C_5^A(Q^2)$ has the form:
\begin{eqnarray}
\label{eq:dipolec5a}
C_5^A(Q^2)=\frac{C^A_5(0)}{\displaystyle \left(1+ \frac{Q^2}{M_{ A\Delta}^2} \right)^2},
\end{eqnarray}
where $M_{A\Delta} = 0.85$~GeV and $C_5^A(0) = 1.10$ as obtained in Ref. \cite{Graczyk:2014dpa}. 

For $F_{\rho}$ is given by~\cite{Hernandez:2007qq}
\begin{equation}
\label{Frho}
F_{\rho}(q)=\frac{1}{\displaystyle 1 - {q^2}/{m_{\rho}^2}},
\end{equation}  
$m_{\rho}=0.7758$~GeV.

\subsection{FN Model}

The $C_3^V$ vector form factor for $N \to \Delta$ transition  reads~\cite{Dufner:1967yj}
\begin{equation}
C_3^V(Q^2) = 2.07 \exp\left(-3.15 \, \frac{\sqrt{Q^2}}{\mathrm{GeV} }\right) \sqrt{ \left(1 + 9 \frac{\sqrt{\displaystyle Q^2}}{\mathrm{GeV}} \right) }.
\end{equation}

The axial form factor $C_5^A(Q^2)$ is parameterized by (\ref{eq:dipolec5a}) with parameters~\cite{Fogli:1979cz}
$M_{A\Delta} = 0.75$~GeV and  $C_5^A(0)=1.18$.
Pion form factor reads
\begin{equation}
\label{Fpi}
F_\pi(q) = \frac{1}{ 1 - q^2/(0.47~\mathrm{GeV}^2) }.
\end{equation}

\section{$\Delta(1232)$ Resonance Widths}	 	
	
\label{Appendix_BreitWigner}

In the HNV model the $\Delta(1232)$ resonance width reads
\begin{eqnarray}
\label{Width_HNV}
\Gamma_{\Delta}^{HNV}(s)&=&\frac{1}{6\pi }\left(\frac{f^{*}}{m_{\pi}} \right)^2 \frac{M}{\sqrt{s}} \nonumber \\ 
& & \!\!\!\!\!\!\!\!\!\!\times \left[\frac{\lambda(s,M^2,m_{\pi}^2)}{2\sqrt{s} }\right]^3\theta(\sqrt{s}-M-m_{\pi}),
\end{eqnarray}
where
\begin{equation}
\lambda(x,y,z)=x^2+y^2+z^2-2 xy - 2 xz - 2 yz.
\end{equation} 

In the FN model the width takes similar  form
\begin{eqnarray}
\label{Width_FN}
\Gamma_{\Delta}^{FN}(s)&=&\frac{1}{6\pi }\left(\frac{f^{*}}{m_{\pi}} \right)^2 \frac{1}{(2\sqrt{s})^5} \left[(\sqrt{s}+M)^2-m_{\pi}^2 \right] \nonumber \\
& & \left[ \lambda(s,M^2,m_{\pi}^2)  \right]^3\theta(\sqrt{s}-M-m_{\pi}).
\end{eqnarray}


\normalem
\bibliographystyle{apsrev4-1}
\bibliography{bibdrat,bibdratbook,bibmoje}

\begin{thebibliography}{57}%
\makeatletter
\providecommand \@ifxundefined [1]{%
 \@ifx{#1\undefined}
}%
\providecommand \@ifnum [1]{%
 \ifnum #1\expandafter \@firstoftwo
 \else \expandafter \@secondoftwo
 \fi
}%
\providecommand \@ifx [1]{%
 \ifx #1\expandafter \@firstoftwo
 \else \expandafter \@secondoftwo
 \fi
}%
\providecommand \natexlab [1]{#1}%
\providecommand \enquote  [1]{``#1''}%
\providecommand \bibnamefont  [1]{#1}%
\providecommand \bibfnamefont [1]{#1}%
\providecommand \citenamefont [1]{#1}%
\providecommand \href@noop [0]{\@secondoftwo}%
\providecommand \href [0]{\begingroup \@sanitize@url \@href}%
\providecommand \@href[1]{\@@startlink{#1}\@@href}%
\providecommand \@@href[1]{\endgroup#1\@@endlink}%
\providecommand \@sanitize@url [0]{\catcode `\\12\catcode `\$12\catcode
  `\&12\catcode `\#12\catcode `\^12\catcode `\_12\catcode `\%12\relax}%
\providecommand \@@startlink[1]{}%
\providecommand \@@endlink[0]{}%
\providecommand \url  [0]{\begingroup\@sanitize@url \@url }%
\providecommand \@url [1]{\endgroup\@href {#1}{\urlprefix }}%
\providecommand \urlprefix  [0]{URL }%
\providecommand \Eprint [0]{\href }%
\providecommand \doibase [0]{http://dx.doi.org/}%
\providecommand \selectlanguage [0]{\@gobble}%
\providecommand \bibinfo  [0]{\@secondoftwo}%
\providecommand \bibfield  [0]{\@secondoftwo}%
\providecommand \translation [1]{[#1]}%
\providecommand \BibitemOpen [0]{}%
\providecommand \bibitemStop [0]{}%
\providecommand \bibitemNoStop [0]{.\EOS\space}%
\providecommand \EOS [0]{\spacefactor3000\relax}%
\providecommand \BibitemShut  [1]{\csname bibitem#1\endcsname}%
\let\auto@bib@innerbib\@empty
\bibitem [{\citenamefont {Llewellyn~Smith}(1972)}]{LlewellynSmith:1971uhs}%
  \BibitemOpen
  \bibfield  {author} {\bibinfo {author} {\bibfnamefont {C.~H.}\ \bibnamefont
  {Llewellyn~Smith}},\ }\bibfield  {booktitle} {\emph {\bibinfo {booktitle}
  {{Gauge Theories and Neutrino Physics, Jacob, 1978:0175}}},\ }\href {\doibase
  10.1016/0370-1573(72)90010-5} {\bibfield  {journal} {\bibinfo  {journal}
  {Phys. Rept.}\ }\textbf {\bibinfo {volume} {3}},\ \bibinfo {pages} {261}
  (\bibinfo {year} {1972})}\BibitemShut {NoStop}%
\bibitem [{\citenamefont {Aguilar-Arevalo}\ \emph {et~al.}(2007)\citenamefont
  {Aguilar-Arevalo} \emph {et~al.}}]{AguilarArevalo:2007it}%
  \BibitemOpen
  \bibfield  {author} {\bibinfo {author} {\bibfnamefont {A.~A.}\ \bibnamefont
  {Aguilar-Arevalo}} \emph {et~al.} (\bibinfo {collaboration} {MiniBooNE}),\
  }\href {\doibase 10.1103/PhysRevLett.98.231801} {\bibfield  {journal}
  {\bibinfo  {journal} {Phys. Rev. Lett.}\ }\textbf {\bibinfo {volume} {98}},\
  \bibinfo {pages} {231801} (\bibinfo {year} {2007})},\ \Eprint
  {http://arxiv.org/abs/0704.1500} {arXiv:0704.1500 [hep-ex]} \BibitemShut
  {NoStop}%
\bibitem [{\citenamefont {Abe}\ \emph {et~al.}(2011)\citenamefont {Abe} \emph
  {et~al.}}]{Abe:2011ks}%
  \BibitemOpen
  \bibfield  {author} {\bibinfo {author} {\bibfnamefont {K.}~\bibnamefont
  {Abe}} \emph {et~al.} (\bibinfo {collaboration} {T2K}),\ }\href {\doibase
  10.1016/j.nima.2011.06.067} {\bibfield  {journal} {\bibinfo  {journal} {Nucl.
  Instrum. Meth.}\ }\textbf {\bibinfo {volume} {A659}},\ \bibinfo {pages} {106}
  (\bibinfo {year} {2011})},\ \Eprint {http://arxiv.org/abs/1106.1238}
  {arXiv:1106.1238 [physics.ins-det]} \BibitemShut {NoStop}%
\bibitem [{\citenamefont {Evans}(2013)}]{Evans:2013pka}%
  \BibitemOpen
  \bibfield  {author} {\bibinfo {author} {\bibfnamefont {J.}~\bibnamefont
  {Evans}} (\bibinfo {collaboration} {MINOS}),\ }\href {\doibase
  10.1155/2013/182537} {\bibfield  {journal} {\bibinfo  {journal} {{Adv. High
  Energy Phys.}}\ }\textbf {\bibinfo {volume} {2013}},\ \bibinfo {pages}
  {182537} (\bibinfo {year} {2013})},\ \Eprint {http://arxiv.org/abs/1307.0721}
  {arXiv:1307.0721 [hep-ex]} \BibitemShut {NoStop}%
\bibitem [{\citenamefont {Ayres}\ \emph {et~al.}(2004)\citenamefont {Ayres}
  \emph {et~al.}}]{Ayres:2004js}%
  \BibitemOpen
  \bibfield  {author} {\bibinfo {author} {\bibfnamefont {D.~S.}\ \bibnamefont
  {Ayres}} \emph {et~al.} (\bibinfo {collaboration} {NOvA}),\ }\href@noop {} {\
   (\bibinfo {year} {2004})},\ \Eprint {http://arxiv.org/abs/hep-ex/0503053}
  {arXiv:hep-ex/0503053 [hep-ex]} \BibitemShut {NoStop}%
\bibitem [{\citenamefont {Aliaga}\ \emph {et~al.}(2014)\citenamefont {Aliaga}
  \emph {et~al.}}]{Aliaga:2013uqz}%
  \BibitemOpen
  \bibfield  {author} {\bibinfo {author} {\bibfnamefont {L.}~\bibnamefont
  {Aliaga}} \emph {et~al.} (\bibinfo {collaboration} {MINERvA}),\ }\href
  {\doibase 10.1016/j.nima.2013.12.053} {\bibfield  {journal} {\bibinfo
  {journal} {Nucl. Instrum. Meth.}\ }\textbf {\bibinfo {volume} {A743}},\
  \bibinfo {pages} {130} (\bibinfo {year} {2014})},\ \Eprint
  {http://arxiv.org/abs/1305.5199} {arXiv:1305.5199 [physics.ins-det]}
  \BibitemShut {NoStop}%
\bibitem [{\citenamefont
  {Mosel}(2016)}]{doi:10.1146/annurev-nucl-102115-044720}%
  \BibitemOpen
  \bibfield  {author} {\bibinfo {author} {\bibfnamefont {U.}~\bibnamefont
  {Mosel}},\ }\href {\doibase 10.1146/annurev-nucl-102115-044720} {\bibfield
  {journal} {\bibinfo  {journal} {{Annual Review of Nuclear and Particle
  Science}}\ }\textbf {\bibinfo {volume} {66}},\ \bibinfo {pages} {171}
  (\bibinfo {year} {2016})}\BibitemShut {NoStop}%
\bibitem [{\citenamefont {Hernandez}\ \emph
  {et~al.}(2007{\natexlab{a}})\citenamefont {Hernandez}, \citenamefont
  {Nieves},\ and\ \citenamefont {Valverde}}]{Hernandez:2007qq}%
  \BibitemOpen
  \bibfield  {author} {\bibinfo {author} {\bibfnamefont {E.}~\bibnamefont
  {Hernandez}}, \bibinfo {author} {\bibfnamefont {J.}~\bibnamefont {Nieves}}, \
  and\ \bibinfo {author} {\bibfnamefont {M.}~\bibnamefont {Valverde}},\ }\href
  {\doibase 10.1103/PhysRevD.76.033005} {\bibfield  {journal} {\bibinfo
  {journal} {Phys. Rev.}\ }\textbf {\bibinfo {volume} {D76}},\ \bibinfo {pages}
  {033005} (\bibinfo {year} {2007}{\natexlab{a}})},\ \Eprint
  {http://arxiv.org/abs/hep-ph/0701149} {arXiv:hep-ph/0701149 [hep-ph]}
  \BibitemShut {NoStop}%
\bibitem [{\citenamefont {Adler}(1968)}]{Adler:1968tw}%
  \BibitemOpen
  \bibfield  {author} {\bibinfo {author} {\bibfnamefont {S.~L.}\ \bibnamefont
  {Adler}},\ }\href {\doibase 10.1016/0003-4916(68)90278-9} {\bibfield
  {journal} {\bibinfo  {journal} {Annals Phys.}\ }\textbf {\bibinfo {volume}
  {50}},\ \bibinfo {pages} {189} (\bibinfo {year} {1968})}\BibitemShut
  {NoStop}%
\bibitem [{\citenamefont {Rein}\ and\ \citenamefont
  {Sehgal}(1981)}]{Rein:1980wg}%
  \BibitemOpen
  \bibfield  {author} {\bibinfo {author} {\bibfnamefont {D.}~\bibnamefont
  {Rein}}\ and\ \bibinfo {author} {\bibfnamefont {L.~M.}\ \bibnamefont
  {Sehgal}},\ }\href {\doibase 10.1016/0003-4916(81)90242-6} {\bibfield
  {journal} {\bibinfo  {journal} {Annals Phys.}\ }\textbf {\bibinfo {volume}
  {133}},\ \bibinfo {pages} {79} (\bibinfo {year} {1981})}\BibitemShut
  {NoStop}%
\bibitem [{\citenamefont {Fogli}\ and\ \citenamefont
  {Nardulli}(1979)}]{Fogli:1979cz}%
  \BibitemOpen
  \bibfield  {author} {\bibinfo {author} {\bibfnamefont {G.~L.}\ \bibnamefont
  {Fogli}}\ and\ \bibinfo {author} {\bibfnamefont {G.}~\bibnamefont
  {Nardulli}},\ }\href {\doibase 10.1016/0550-3213(79)90233-5} {\bibfield
  {journal} {\bibinfo  {journal} {Nucl. Phys.}\ }\textbf {\bibinfo {volume}
  {B160}},\ \bibinfo {pages} {116} (\bibinfo {year} {1979})}\BibitemShut
  {NoStop}%
\bibitem [{\citenamefont {Rein}(1987)}]{Rein:1987cb}%
  \BibitemOpen
  \bibfield  {author} {\bibinfo {author} {\bibfnamefont {D.}~\bibnamefont
  {Rein}},\ }\href {\doibase 10.1007/BF01561054} {\bibfield  {journal}
  {\bibinfo  {journal} {Z. Phys.}\ }\textbf {\bibinfo {volume} {C35}},\
  \bibinfo {pages} {43} (\bibinfo {year} {1987})}\BibitemShut {NoStop}%
\bibitem [{\citenamefont {Hernandez}\ \emph
  {et~al.}(2007{\natexlab{b}})\citenamefont {Hernandez}, \citenamefont
  {Nieves},\ and\ \citenamefont {Valverde}}]{Hernandez:2006yg}%
  \BibitemOpen
  \bibfield  {author} {\bibinfo {author} {\bibfnamefont {E.}~\bibnamefont
  {Hernandez}}, \bibinfo {author} {\bibfnamefont {J.}~\bibnamefont {Nieves}}, \
  and\ \bibinfo {author} {\bibfnamefont {M.}~\bibnamefont {Valverde}},\ }\href
  {\doibase 10.1016/j.physletb.2007.02.051} {\bibfield  {journal} {\bibinfo
  {journal} {Phys. Lett.}\ }\textbf {\bibinfo {volume} {B647}},\ \bibinfo
  {pages} {452} (\bibinfo {year} {2007}{\natexlab{b}})},\ \Eprint
  {http://arxiv.org/abs/hep-ph/0608119} {arXiv:hep-ph/0608119 [hep-ph]}
  \BibitemShut {NoStop}%
\bibitem [{\citenamefont {Nakamura}\ \emph {et~al.}(2015)\citenamefont
  {Nakamura}, \citenamefont {Kamano},\ and\ \citenamefont
  {Sato}}]{Nakamura:2015rta}%
  \BibitemOpen
  \bibfield  {author} {\bibinfo {author} {\bibfnamefont {S.~X.}\ \bibnamefont
  {Nakamura}}, \bibinfo {author} {\bibfnamefont {H.}~\bibnamefont {Kamano}}, \
  and\ \bibinfo {author} {\bibfnamefont {T.}~\bibnamefont {Sato}},\ }\href
  {\doibase 10.1103/PhysRevD.92.074024} {\bibfield  {journal} {\bibinfo
  {journal} {Phys. Rev.}\ }\textbf {\bibinfo {volume} {D92}},\ \bibinfo {pages}
  {074024} (\bibinfo {year} {2015})},\ \Eprint
  {http://arxiv.org/abs/1506.03403} {arXiv:1506.03403 [hep-ph]} \BibitemShut
  {NoStop}%
\bibitem [{\citenamefont {Serot}\ and\ \citenamefont
  {Zhang}(2012)}]{Serot:2012rd}%
  \BibitemOpen
  \bibfield  {author} {\bibinfo {author} {\bibfnamefont {B.~D.}\ \bibnamefont
  {Serot}}\ and\ \bibinfo {author} {\bibfnamefont {X.}~\bibnamefont {Zhang}},\
  }\href {\doibase 10.1103/PhysRevC.86.015501} {\bibfield  {journal} {\bibinfo
  {journal} {Phys. Rev.}\ }\textbf {\bibinfo {volume} {C86}},\ \bibinfo {pages}
  {015501} (\bibinfo {year} {2012})},\ \Eprint {http://arxiv.org/abs/1206.3812}
  {arXiv:1206.3812 [nucl-th]} \BibitemShut {NoStop}%
\bibitem [{\citenamefont {Lalakulich}\ \emph {et~al.}(2010)\citenamefont
  {Lalakulich}, \citenamefont {Leitner}, \citenamefont {Buss},\ and\
  \citenamefont {Mosel}}]{Lalakulich:2010ss}%
  \BibitemOpen
  \bibfield  {author} {\bibinfo {author} {\bibfnamefont {O.}~\bibnamefont
  {Lalakulich}}, \bibinfo {author} {\bibfnamefont {T.}~\bibnamefont {Leitner}},
  \bibinfo {author} {\bibfnamefont {O.}~\bibnamefont {Buss}}, \ and\ \bibinfo
  {author} {\bibfnamefont {U.}~\bibnamefont {Mosel}},\ }\href {\doibase
  10.1103/PhysRevD.82.093001} {\bibfield  {journal} {\bibinfo  {journal} {Phys.
  Rev.}\ }\textbf {\bibinfo {volume} {D82}},\ \bibinfo {pages} {093001}
  (\bibinfo {year} {2010})},\ \Eprint {http://arxiv.org/abs/1007.0925}
  {arXiv:1007.0925 [hep-ph]} \BibitemShut {NoStop}%
\bibitem [{\citenamefont {Leitner}\ \emph {et~al.}(2009)\citenamefont
  {Leitner}, \citenamefont {Buss}, \citenamefont {Alvarez-Ruso},\ and\
  \citenamefont {Mosel}}]{Leitner:2008ue}%
  \BibitemOpen
  \bibfield  {author} {\bibinfo {author} {\bibfnamefont {T.}~\bibnamefont
  {Leitner}}, \bibinfo {author} {\bibfnamefont {O.}~\bibnamefont {Buss}},
  \bibinfo {author} {\bibfnamefont {L.}~\bibnamefont {Alvarez-Ruso}}, \ and\
  \bibinfo {author} {\bibfnamefont {U.}~\bibnamefont {Mosel}},\ }\href
  {\doibase 10.1103/PhysRevC.79.034601} {\bibfield  {journal} {\bibinfo
  {journal} {Phys. Rev.}\ }\textbf {\bibinfo {volume} {C79}},\ \bibinfo {pages}
  {034601} (\bibinfo {year} {2009})},\ \Eprint {http://arxiv.org/abs/0812.0587}
  {arXiv:0812.0587 [nucl-th]} \BibitemShut {NoStop}%
\bibitem [{\citenamefont {Rafi~Alam}\ \emph {et~al.}(2016)\citenamefont
  {Rafi~Alam}, \citenamefont {Sajjad~Athar}, \citenamefont {Chauhan},\ and\
  \citenamefont {Singh}}]{Alam:2015gaa}%
  \BibitemOpen
  \bibfield  {author} {\bibinfo {author} {\bibfnamefont {M.}~\bibnamefont
  {Rafi~Alam}}, \bibinfo {author} {\bibfnamefont {M.}~\bibnamefont
  {Sajjad~Athar}}, \bibinfo {author} {\bibfnamefont {S.}~\bibnamefont
  {Chauhan}}, \ and\ \bibinfo {author} {\bibfnamefont {S.~K.}\ \bibnamefont
  {Singh}},\ }\href {\doibase 10.1142/S0218301316500105} {\bibfield  {journal}
  {\bibinfo  {journal} {Int. J. Mod. Phys.}\ }\textbf {\bibinfo {volume}
  {E25}},\ \bibinfo {pages} {1650010} (\bibinfo {year} {2016})},\ \Eprint
  {http://arxiv.org/abs/1509.08622} {arXiv:1509.08622 [hep-ph]} \BibitemShut
  {NoStop}%
\bibitem [{\citenamefont {Barbero}\ \emph {et~al.}(2008)\citenamefont
  {Barbero}, \citenamefont {Lopez~Castro},\ and\ \citenamefont
  {Mariano}}]{Barbero:2008zza}%
  \BibitemOpen
  \bibfield  {author} {\bibinfo {author} {\bibfnamefont {C.}~\bibnamefont
  {Barbero}}, \bibinfo {author} {\bibfnamefont {G.}~\bibnamefont
  {Lopez~Castro}}, \ and\ \bibinfo {author} {\bibfnamefont {A.}~\bibnamefont
  {Mariano}},\ }\href {\doibase 10.1016/j.physletb.2008.05.011} {\bibfield
  {journal} {\bibinfo  {journal} {Phys. Lett.}\ }\textbf {\bibinfo {volume}
  {B664}},\ \bibinfo {pages} {70} (\bibinfo {year} {2008})}\BibitemShut
  {NoStop}%
\bibitem [{\citenamefont {Gonzalez-Jimenez}\ \emph {et~al.}(2017)\citenamefont
  {Gonzalez-Jimenez}, \citenamefont {Jachowicz}, \citenamefont {Niewczas},
  \citenamefont {Nys}, \citenamefont {Pandey}, \citenamefont {Van~Cuyck},\ and\
  \citenamefont {Van~Dessel}}]{Gonzalez-Jimenez:2016qqq}%
  \BibitemOpen
  \bibfield  {author} {\bibinfo {author} {\bibfnamefont {R.}~\bibnamefont
  {Gonzalez-Jimenez}}, \bibinfo {author} {\bibfnamefont {N.}~\bibnamefont
  {Jachowicz}}, \bibinfo {author} {\bibfnamefont {K.}~\bibnamefont {Niewczas}},
  \bibinfo {author} {\bibfnamefont {J.}~\bibnamefont {Nys}}, \bibinfo {author}
  {\bibfnamefont {V.}~\bibnamefont {Pandey}}, \bibinfo {author} {\bibfnamefont
  {T.}~\bibnamefont {Van~Cuyck}}, \ and\ \bibinfo {author} {\bibfnamefont
  {N.}~\bibnamefont {Van~Dessel}},\ }\href {\doibase
  10.1103/PhysRevD.95.113007} {\bibfield  {journal} {\bibinfo  {journal} {Phys.
  Rev.}\ }\textbf {\bibinfo {volume} {D95}},\ \bibinfo {pages} {113007}
  (\bibinfo {year} {2017})},\ \Eprint {http://arxiv.org/abs/1612.05511}
  {arXiv:1612.05511 [nucl-th]} \BibitemShut {NoStop}%
\bibitem [{\citenamefont {Alvarez-Ruso}\ \emph {et~al.}(2014)\citenamefont
  {Alvarez-Ruso}, \citenamefont {Hayato},\ and\ \citenamefont
  {Nieves}}]{Alvarez-Ruso:2014bla}%
  \BibitemOpen
  \bibfield  {author} {\bibinfo {author} {\bibfnamefont {L.}~\bibnamefont
  {Alvarez-Ruso}}, \bibinfo {author} {\bibfnamefont {Y.}~\bibnamefont
  {Hayato}}, \ and\ \bibinfo {author} {\bibfnamefont {J.}~\bibnamefont
  {Nieves}},\ }\href {\doibase 10.1088/1367-2630/16/7/075015} {\bibfield
  {journal} {\bibinfo  {journal} {New J. Phys.}\ }\textbf {\bibinfo {volume}
  {16}},\ \bibinfo {pages} {075015} (\bibinfo {year} {2014})},\ \Eprint
  {http://arxiv.org/abs/1403.2673} {arXiv:1403.2673 [hep-ph]} \BibitemShut
  {NoStop}%
\bibitem [{\citenamefont {Radecky}\ \emph {et~al.}(1982)\citenamefont
  {Radecky}, \citenamefont {Barnes}, \citenamefont {Carmony}, \citenamefont
  {Garfinkel}, \citenamefont {Derrick}, \citenamefont {Fernandez},
  \citenamefont {Hyman},\ and\ \citenamefont {Levman~{\it et
  al.}}}]{Radecky:1981fn}%
  \BibitemOpen
  \bibfield  {author} {\bibinfo {author} {\bibfnamefont {G.~M.}\ \bibnamefont
  {Radecky}}, \bibinfo {author} {\bibfnamefont {V.~E.}\ \bibnamefont {Barnes}},
  \bibinfo {author} {\bibfnamefont {D.~D.}\ \bibnamefont {Carmony}}, \bibinfo
  {author} {\bibfnamefont {A.~F.}\ \bibnamefont {Garfinkel}}, \bibinfo {author}
  {\bibfnamefont {M.}~\bibnamefont {Derrick}}, \bibinfo {author} {\bibfnamefont
  {E.}~\bibnamefont {Fernandez}}, \bibinfo {author} {\bibfnamefont
  {L.}~\bibnamefont {Hyman}}, \ and\ \bibinfo {author} {\bibfnamefont
  {G.}~\bibnamefont {Levman~{\it et al.}}},\ }\href@noop {} {\bibfield
  {journal} {\bibinfo  {journal} {Phys.\ Rev.\ D}\ }\textbf {\bibinfo {volume}
  {{\bf 26}}},\ \bibinfo {pages} {3297} (\bibinfo {year} {1982})},\ \bibinfo
  {note} {[Erratum-ibid.\ D {\bf 26} (1982) 3297]}\BibitemShut {NoStop}%
\bibitem [{\citenamefont {Kitagaki}\ \emph {et~al.}(1986)\citenamefont
  {Kitagaki}, \citenamefont {Yuta}, \citenamefont {Tanaka}, \citenamefont
  {Yamaguchi}, \citenamefont {Abe} \emph {et~al.}}]{Kitagaki:1986ct}%
  \BibitemOpen
  \bibfield  {author} {\bibinfo {author} {\bibfnamefont {T.}~\bibnamefont
  {Kitagaki}}, \bibinfo {author} {\bibfnamefont {H.}~\bibnamefont {Yuta}},
  \bibinfo {author} {\bibfnamefont {S.}~\bibnamefont {Tanaka}}, \bibinfo
  {author} {\bibfnamefont {A.}~\bibnamefont {Yamaguchi}}, \bibinfo {author}
  {\bibfnamefont {K.}~\bibnamefont {Abe}},  \emph {et~al.},\ }\href {\doibase
  10.1103/PhysRevD.34.2554} {\bibfield  {journal} {\bibinfo  {journal} {Phys.
  Rev. D}\ }\textbf {\bibinfo {volume} {34}},\ \bibinfo {pages} {2554}
  (\bibinfo {year} {1986})}\BibitemShut {NoStop}%
\bibitem [{\citenamefont {Altinok}\ \emph {et~al.}(2017)\citenamefont {Altinok}
  \emph {et~al.}}]{Altinok:2017xua}%
  \BibitemOpen
  \bibfield  {author} {\bibinfo {author} {\bibfnamefont {O.}~\bibnamefont
  {Altinok}} \emph {et~al.} (\bibinfo {collaboration} {MINERvA}),\ }\href@noop
  {} {\  (\bibinfo {year} {2017})},\ \Eprint {http://arxiv.org/abs/1708.03723}
  {arXiv:1708.03723 [hep-ex]} \BibitemShut {NoStop}%
\bibitem [{\citenamefont {Akhiezer}\ and\ \citenamefont
  {Rekalo}(1968)}]{Akhiezer:1968ek}%
  \BibitemOpen
  \bibfield  {author} {\bibinfo {author} {\bibfnamefont {A.~I.}\ \bibnamefont
  {Akhiezer}}\ and\ \bibinfo {author} {\bibfnamefont {M.}~\bibnamefont
  {Rekalo}},\ }\href@noop {} {\bibfield  {journal} {\bibinfo  {journal} {Sov.
  Phys. Dokl.}\ }\textbf {\bibinfo {volume} {13}},\ \bibinfo {pages} {572}
  (\bibinfo {year} {1968})},\ \bibinfo {note} {[Dokl. Akad. Nauk Ser.
  Fiz.180,1081(1968)]}\BibitemShut {NoStop}%
\bibitem [{\citenamefont {Akhiezer}\ and\ \citenamefont
  {Rekalo}(1974)}]{Akhiezer:1974em}%
  \BibitemOpen
  \bibfield  {author} {\bibinfo {author} {\bibfnamefont {A.~I.}\ \bibnamefont
  {Akhiezer}}\ and\ \bibinfo {author} {\bibfnamefont {M.}~\bibnamefont
  {Rekalo}},\ }\href@noop {} {\bibfield  {journal} {\bibinfo  {journal} {Sov.
  J. Part. Nucl.}\ }\textbf {\bibinfo {volume} {4}},\ \bibinfo {pages} {277}
  (\bibinfo {year} {1974})},\ \bibinfo {note} {[Fiz. Elem. Chast. Atom.
  Yadra4,662(1973)]}\BibitemShut {NoStop}%
\bibitem [{\citenamefont {Ohlsen}(1972)}]{Ohlsen:1972zz}%
  \BibitemOpen
  \bibfield  {author} {\bibinfo {author} {\bibfnamefont {G.~G.}\ \bibnamefont
  {Ohlsen}},\ }\href {\doibase 10.1088/0034-4885/35/2/305} {\bibfield
  {journal} {\bibinfo  {journal} {Rept. Prog. Phys.}\ }\textbf {\bibinfo
  {volume} {35}},\ \bibinfo {pages} {717} (\bibinfo {year} {1972})}\BibitemShut
  {NoStop}%
\bibitem [{\citenamefont {Arnold}\ \emph {et~al.}(1981)\citenamefont {Arnold},
  \citenamefont {Carlson},\ and\ \citenamefont {Gross}}]{Arnold:1980zj}%
  \BibitemOpen
  \bibfield  {author} {\bibinfo {author} {\bibfnamefont {R.~G.}\ \bibnamefont
  {Arnold}}, \bibinfo {author} {\bibfnamefont {C.~E.}\ \bibnamefont {Carlson}},
  \ and\ \bibinfo {author} {\bibfnamefont {F.}~\bibnamefont {Gross}},\ }\href
  {\doibase 10.1103/PhysRevC.23.363} {\bibfield  {journal} {\bibinfo  {journal}
  {Phys. Rev.}\ }\textbf {\bibinfo {volume} {C23}},\ \bibinfo {pages} {363}
  (\bibinfo {year} {1981})}\BibitemShut {NoStop}%
\bibitem [{\citenamefont {Donnelly}\ and\ \citenamefont
  {Raskin}(1986)}]{Donnelly:1985ry}%
  \BibitemOpen
  \bibfield  {author} {\bibinfo {author} {\bibfnamefont {T.~W.}\ \bibnamefont
  {Donnelly}}\ and\ \bibinfo {author} {\bibfnamefont {A.~S.}\ \bibnamefont
  {Raskin}},\ }\href {\doibase 10.1016/0003-4916(86)90173-9} {\bibfield
  {journal} {\bibinfo  {journal} {Annals Phys.}\ }\textbf {\bibinfo {volume}
  {169}},\ \bibinfo {pages} {247} (\bibinfo {year} {1986})}\BibitemShut
  {NoStop}%
\bibitem [{\citenamefont {Afanasev}\ \emph {et~al.}(2017)\citenamefont
  {Afanasev}, \citenamefont {Blunden}, \citenamefont {Hasell},\ and\
  \citenamefont {Raue}}]{Afanasev:2017gsk}%
  \BibitemOpen
  \bibfield  {author} {\bibinfo {author} {\bibfnamefont {A.}~\bibnamefont
  {Afanasev}}, \bibinfo {author} {\bibfnamefont {P.~G.}\ \bibnamefont
  {Blunden}}, \bibinfo {author} {\bibfnamefont {D.}~\bibnamefont {Hasell}}, \
  and\ \bibinfo {author} {\bibfnamefont {B.~A.}\ \bibnamefont {Raue}},\ }\href
  {\doibase 10.1016/j.ppnp.2017.03.004} {\bibfield  {journal} {\bibinfo
  {journal} {Prog. Part. Nucl. Phys.}\ }\textbf {\bibinfo {volume} {95}},\
  \bibinfo {pages} {245} (\bibinfo {year} {2017})},\ \Eprint
  {http://arxiv.org/abs/1703.03874} {arXiv:1703.03874 [nucl-ex]} \BibitemShut
  {NoStop}%
\bibitem [{\citenamefont {Bilenky}\ and\ \citenamefont
  {Christova}(2013{\natexlab{a}})}]{Bilenky:2013iua}%
  \BibitemOpen
  \bibfield  {author} {\bibinfo {author} {\bibfnamefont {S.~M.}\ \bibnamefont
  {Bilenky}}\ and\ \bibinfo {author} {\bibfnamefont {E.}~\bibnamefont
  {Christova}},\ }\href {\doibase 10.1134/S154747711307011X} {\bibfield
  {journal} {\bibinfo  {journal} {Phys. Part. Nucl. Lett.}\ }\textbf {\bibinfo
  {volume} {10}},\ \bibinfo {pages} {651} (\bibinfo {year}
  {2013}{\natexlab{a}})},\ \Eprint {http://arxiv.org/abs/1307.7275}
  {arXiv:1307.7275 [hep-ph]} \BibitemShut {NoStop}%
\bibitem [{\citenamefont {Bilenky}\ and\ \citenamefont
  {Christova}(2013{\natexlab{b}})}]{Bilenky:2013fra}%
  \BibitemOpen
  \bibfield  {author} {\bibinfo {author} {\bibfnamefont {S.~M.}\ \bibnamefont
  {Bilenky}}\ and\ \bibinfo {author} {\bibfnamefont {E.}~\bibnamefont
  {Christova}},\ }\href {\doibase 10.1088/0954-3899/40/7/075004} {\bibfield
  {journal} {\bibinfo  {journal} {J. Phys.}\ }\textbf {\bibinfo {volume}
  {G40}},\ \bibinfo {pages} {075004} (\bibinfo {year} {2013}{\natexlab{b}})},\
  \Eprint {http://arxiv.org/abs/1303.3710} {arXiv:1303.3710 [hep-ph]}
  \BibitemShut {NoStop}%
\bibitem [{\citenamefont {Akbar}\ \emph {et~al.}(2017)\citenamefont {Akbar},
  \citenamefont {Sajjad~Athar}, \citenamefont {Fatima},\ and\ \citenamefont
  {Singh}}]{Akbar:2017qsf}%
  \BibitemOpen
  \bibfield  {author} {\bibinfo {author} {\bibfnamefont {F.}~\bibnamefont
  {Akbar}}, \bibinfo {author} {\bibfnamefont {M.}~\bibnamefont {Sajjad~Athar}},
  \bibinfo {author} {\bibfnamefont {A.}~\bibnamefont {Fatima}}, \ and\ \bibinfo
  {author} {\bibfnamefont {S.~K.}\ \bibnamefont {Singh}},\ }\href {\doibase
  10.1140/epja/i2017-12340-4} {\bibfield  {journal} {\bibinfo  {journal} {Eur.
  Phys. J.}\ }\textbf {\bibinfo {volume} {A53}},\ \bibinfo {pages} {154}
  (\bibinfo {year} {2017})},\ \Eprint {http://arxiv.org/abs/1704.04580}
  {arXiv:1704.04580 [hep-ph]} \BibitemShut {NoStop}%
\bibitem [{\citenamefont {Akbar}\ \emph {et~al.}(2016)\citenamefont {Akbar},
  \citenamefont {Rafi~Alam}, \citenamefont {Sajjad~Athar},\ and\ \citenamefont
  {Singh}}]{Akbar:2016awk}%
  \BibitemOpen
  \bibfield  {author} {\bibinfo {author} {\bibfnamefont {F.}~\bibnamefont
  {Akbar}}, \bibinfo {author} {\bibfnamefont {M.}~\bibnamefont {Rafi~Alam}},
  \bibinfo {author} {\bibfnamefont {M.}~\bibnamefont {Sajjad~Athar}}, \ and\
  \bibinfo {author} {\bibfnamefont {S.~K.}\ \bibnamefont {Singh}},\ }\href
  {\doibase 10.1103/PhysRevD.94.114031} {\bibfield  {journal} {\bibinfo
  {journal} {Phys. Rev.}\ }\textbf {\bibinfo {volume} {D94}},\ \bibinfo {pages}
  {114031} (\bibinfo {year} {2016})},\ \Eprint
  {http://arxiv.org/abs/1608.02103} {arXiv:1608.02103 [hep-ph]} \BibitemShut
  {NoStop}%
\bibitem [{\citenamefont {Kuzmin}\ \emph {et~al.}(2004)\citenamefont {Kuzmin},
  \citenamefont {Lyubushkin},\ and\ \citenamefont {Naumov}}]{Kuzmin:2003ji}%
  \BibitemOpen
  \bibfield  {author} {\bibinfo {author} {\bibfnamefont {K.~S.}\ \bibnamefont
  {Kuzmin}}, \bibinfo {author} {\bibfnamefont {V.~V.}\ \bibnamefont
  {Lyubushkin}}, \ and\ \bibinfo {author} {\bibfnamefont {V.~A.}\ \bibnamefont
  {Naumov}},\ }\bibfield  {booktitle} {\emph {\bibinfo {booktitle}
  {{Proceedings, 10th Advanced Research Workshop on High-Energy Spin Physics
  (SPIN-03): Dubna, Russia, September 16-20, 2003}}},\ }\href {\doibase
  10.1142/S0217732304016172} {\bibfield  {journal} {\bibinfo  {journal} {Mod.
  Phys. Lett.}\ }\textbf {\bibinfo {volume} {A19}},\ \bibinfo {pages} {2815}
  (\bibinfo {year} {2004})},\ \bibinfo {note} {[,125(2003)]},\ \Eprint
  {http://arxiv.org/abs/hep-ph/0312107} {arXiv:hep-ph/0312107 [hep-ph]}
  \BibitemShut {NoStop}%
\bibitem [{\citenamefont {Hagiwara}\ \emph {et~al.}(2003)\citenamefont
  {Hagiwara}, \citenamefont {Mawatari},\ and\ \citenamefont
  {Yokoya}}]{Hagiwara:2003di}%
  \BibitemOpen
  \bibfield  {author} {\bibinfo {author} {\bibfnamefont {K.}~\bibnamefont
  {Hagiwara}}, \bibinfo {author} {\bibfnamefont {K.}~\bibnamefont {Mawatari}},
  \ and\ \bibinfo {author} {\bibfnamefont {H.}~\bibnamefont {Yokoya}},\ }\href
  {\doibase 10.1016/S0550-3213(03)00575-3} {\bibfield  {journal} {\bibinfo
  {journal} {Nucl. Phys.}\ }\textbf {\bibinfo {volume} {B668}},\ \bibinfo
  {pages} {364} (\bibinfo {year} {2003})},\ \bibinfo {note} {[Erratum: Nucl.
  Phys.B701,405(2004)]},\ \Eprint {http://arxiv.org/abs/hep-ph/0305324}
  {arXiv:hep-ph/0305324 [hep-ph]} \BibitemShut {NoStop}%
\bibitem [{\citenamefont {Kuzmin}\ \emph {et~al.}(2005)\citenamefont {Kuzmin},
  \citenamefont {Lyubushkin},\ and\ \citenamefont {Naumov}}]{Kuzmin:2004yb}%
  \BibitemOpen
  \bibfield  {author} {\bibinfo {author} {\bibfnamefont {K.~S.}\ \bibnamefont
  {Kuzmin}}, \bibinfo {author} {\bibfnamefont {V.~V.}\ \bibnamefont
  {Lyubushkin}}, \ and\ \bibinfo {author} {\bibfnamefont {V.~A.}\ \bibnamefont
  {Naumov}},\ }\bibfield  {booktitle} {\emph {\bibinfo {booktitle}
  {{Proceedings, 3rd International Workshop on Neutrino-nucleus interactions in
  the few GeV region (NUINT 04): Assergi, Italy, March 17-21, 2004}}},\ }\href
  {\doibase 10.1016/j.nuclphysbps.2004.11.221} {\bibfield  {journal} {\bibinfo
  {journal} {Nucl. Phys. Proc. Suppl.}\ }\textbf {\bibinfo {volume} {139}},\
  \bibinfo {pages} {154} (\bibinfo {year} {2005})},\ \bibinfo {note}
  {[,154(2004)]},\ \Eprint {http://arxiv.org/abs/hep-ph/0408107}
  {arXiv:hep-ph/0408107 [hep-ph]} \BibitemShut {NoStop}%
\bibitem [{\citenamefont {Graczyk}(2005)}]{Graczyk:2004uy}%
  \BibitemOpen
  \bibfield  {author} {\bibinfo {author} {\bibfnamefont {K.~M.}\ \bibnamefont
  {Graczyk}},\ }\href {\doibase 10.1016/j.nuclphysa.2004.10.029} {\bibfield
  {journal} {\bibinfo  {journal} {Nucl. Phys.}\ }\textbf {\bibinfo {volume}
  {A748}},\ \bibinfo {pages} {313} (\bibinfo {year} {2005})},\ \Eprint
  {http://arxiv.org/abs/hep-ph/0407275} {arXiv:hep-ph/0407275 [hep-ph]}
  \BibitemShut {NoStop}%
\bibitem [{\citenamefont {Dombey}(1969)}]{Dombey:1969wk}%
  \BibitemOpen
  \bibfield  {author} {\bibinfo {author} {\bibfnamefont {N.}~\bibnamefont
  {Dombey}},\ }\href {\doibase 10.1103/RevModPhys.41.236} {\bibfield  {journal}
  {\bibinfo  {journal} {Rev. Mod. Phys.}\ }\textbf {\bibinfo {volume} {41}},\
  \bibinfo {pages} {236} (\bibinfo {year} {1969})}\BibitemShut {NoStop}%
\bibitem [{\citenamefont {{W. Greiner}}(1992)}]{Greiner__QED_book}%
  \BibitemOpen
  \bibfield  {author} {\bibinfo {author} {\bibnamefont {{W. Greiner}}},\ }\href
  {http://www.springer.com/us/book/9783540875604} {\emph {\bibinfo {title}
  {{Quantum Electrodynamics}}}}\ (\bibinfo  {publisher} {Springer-Verlag Berlin
  Heidelberg},\ \bibinfo {year} {1992})\BibitemShut {NoStop}%
\bibitem [{\citenamefont {Maximon}\ and\ \citenamefont
  {Parke}(2000)}]{Maximon:2000hk}%
  \BibitemOpen
  \bibfield  {author} {\bibinfo {author} {\bibfnamefont {L.~C.}\ \bibnamefont
  {Maximon}}\ and\ \bibinfo {author} {\bibfnamefont {W.~C.}\ \bibnamefont
  {Parke}},\ }\href {\doibase 10.1103/PhysRevC.61.045502} {\bibfield  {journal}
  {\bibinfo  {journal} {Phys. Rev.}\ }\textbf {\bibinfo {volume} {C61}},\
  \bibinfo {pages} {045502} (\bibinfo {year} {2000})},\ \Eprint
  {http://arxiv.org/abs/nucl-th/0002057} {arXiv:nucl-th/0002057 [nucl-th]}
  \BibitemShut {NoStop}%
\bibitem [{\citenamefont {{Walecka J.
  D.}}()}]{Walecka_electron_scattering_textbook}%
  \BibitemOpen
  \bibfield  {author} {\bibinfo {author} {\bibnamefont {{Walecka J. D.}}},\
  }\href@noop {} {\emph {\bibinfo {title} {{Electron Scattering for Nuclear and
  Nucleon Structure}}}}\BibitemShut {NoStop}%
\bibitem [{\citenamefont {Alvarez-Ruso}\ \emph {et~al.}(2016)\citenamefont
  {Alvarez-Ruso}, \citenamefont {Hernandez}, \citenamefont {Nieves},\ and\
  \citenamefont {Vicente~Vacas}}]{Alvarez-Ruso:2015eva}%
  \BibitemOpen
  \bibfield  {author} {\bibinfo {author} {\bibfnamefont {L.}~\bibnamefont
  {Alvarez-Ruso}}, \bibinfo {author} {\bibfnamefont {E.}~\bibnamefont
  {Hernandez}}, \bibinfo {author} {\bibfnamefont {J.}~\bibnamefont {Nieves}}, \
  and\ \bibinfo {author} {\bibfnamefont {M.~J.}\ \bibnamefont
  {Vicente~Vacas}},\ }\href {\doibase 10.1103/PhysRevD.93.014016} {\bibfield
  {journal} {\bibinfo  {journal} {Phys. Rev.}\ }\textbf {\bibinfo {volume}
  {D93}},\ \bibinfo {pages} {014016} (\bibinfo {year} {2016})},\ \Eprint
  {http://arxiv.org/abs/1510.06266} {arXiv:1510.06266 [hep-ph]} \BibitemShut
  {NoStop}%
\bibitem [{\citenamefont {Hernandez}\ and\ \citenamefont
  {Nieves}(2017)}]{Hernandez:2016yfb}%
  \BibitemOpen
  \bibfield  {author} {\bibinfo {author} {\bibfnamefont {E.}~\bibnamefont
  {Hernandez}}\ and\ \bibinfo {author} {\bibfnamefont {J.}~\bibnamefont
  {Nieves}},\ }\href {\doibase 10.1103/PhysRevD.95.053007} {\bibfield
  {journal} {\bibinfo  {journal} {Phys. Rev.}\ }\textbf {\bibinfo {volume}
  {D95}},\ \bibinfo {pages} {053007} (\bibinfo {year} {2017})},\ \Eprint
  {http://arxiv.org/abs/1612.02343} {arXiv:1612.02343 [hep-ph]} \BibitemShut
  {NoStop}%
\bibitem [{\citenamefont {Rarita}\ and\ \citenamefont
  {Schwinger}(1941)}]{Rarita:1941mf}%
  \BibitemOpen
  \bibfield  {author} {\bibinfo {author} {\bibfnamefont {W.}~\bibnamefont
  {Rarita}}\ and\ \bibinfo {author} {\bibfnamefont {J.}~\bibnamefont
  {Schwinger}},\ }\href {\doibase 10.1103/PhysRev.60.61} {\bibfield  {journal}
  {\bibinfo  {journal} {Phys. Rev.}\ }\textbf {\bibinfo {volume} {60}},\
  \bibinfo {pages} {61} (\bibinfo {year} {1941})}\BibitemShut {NoStop}%
\bibitem [{\citenamefont {Jones}\ and\ \citenamefont
  {Scadron}(1973)}]{Jones:1972ky}%
  \BibitemOpen
  \bibfield  {author} {\bibinfo {author} {\bibfnamefont {H.}~\bibnamefont
  {Jones}}\ and\ \bibinfo {author} {\bibfnamefont {M.}~\bibnamefont
  {Scadron}},\ }\href {\doibase 10.1016/0003-4916(73)90476-4} {\bibfield
  {journal} {\bibinfo  {journal} {Annals Phys.}\ }\textbf {\bibinfo {volume}
  {81}},\ \bibinfo {pages} {1} (\bibinfo {year} {1973})}\BibitemShut {NoStop}%
\bibitem [{\citenamefont {Graczyk}\ \emph {et~al.}(2014)\citenamefont
  {Graczyk}, \citenamefont {Zmuda},\ and\ \citenamefont
  {Sobczyk}}]{Graczyk:2014dpa}%
  \BibitemOpen
  \bibfield  {author} {\bibinfo {author} {\bibfnamefont {K.~M.}\ \bibnamefont
  {Graczyk}}, \bibinfo {author} {\bibfnamefont {J.}~\bibnamefont {Zmuda}}, \
  and\ \bibinfo {author} {\bibfnamefont {J.~T.}\ \bibnamefont {Sobczyk}},\
  }\href {\doibase 10.1103/PhysRevD.90.093001} {\bibfield  {journal} {\bibinfo
  {journal} {Phys. Rev.}\ }\textbf {\bibinfo {volume} {D90}},\ \bibinfo {pages}
  {093001} (\bibinfo {year} {2014})},\ \Eprint {http://arxiv.org/abs/1407.5445}
  {arXiv:1407.5445 [hep-ph]} \BibitemShut {NoStop}%
\bibitem [{\citenamefont {Schreiner}\ and\ \citenamefont
  {Von~Hippel}(1973)}]{Schreiner:1973mj}%
  \BibitemOpen
  \bibfield  {author} {\bibinfo {author} {\bibfnamefont {P.~A.}\ \bibnamefont
  {Schreiner}}\ and\ \bibinfo {author} {\bibfnamefont {F.}~\bibnamefont
  {Von~Hippel}},\ }\href {\doibase 10.1016/0550-3213(73)90588-9} {\bibfield
  {journal} {\bibinfo  {journal} {Nucl. Phys.}\ }\textbf {\bibinfo {volume}
  {B58}},\ \bibinfo {pages} {333} (\bibinfo {year} {1973})}\BibitemShut
  {NoStop}%
\bibitem [{\citenamefont {Graczyk}(2009)}]{Graczyk:2009zh}%
  \BibitemOpen
  \bibfield  {author} {\bibinfo {author} {\bibfnamefont {K.~M.}\ \bibnamefont
  {Graczyk}},\ }\bibfield  {booktitle} {\emph {\bibinfo {booktitle}
  {{Proceedings, Europhysics Conference on High energy physics (EPS-HEP 2009):
  Cracow, Poland, July 16-22, 2009}}},\ }\href@noop {} {\bibfield  {journal}
  {\bibinfo  {journal} {PoS}\ }\textbf {\bibinfo {volume} {EPS-HEP2009}},\
  \bibinfo {pages} {286} (\bibinfo {year} {2009})},\ \Eprint
  {http://arxiv.org/abs/0909.5084} {arXiv:0909.5084 [hep-ph]} \BibitemShut
  {NoStop}%
\bibitem [{\citenamefont {Graczyk}\ and\ \citenamefont
  {Sobczyk}(2008)}]{Graczyk:2007bc}%
  \BibitemOpen
  \bibfield  {author} {\bibinfo {author} {\bibfnamefont {K.~M.}\ \bibnamefont
  {Graczyk}}\ and\ \bibinfo {author} {\bibfnamefont {J.~T.}\ \bibnamefont
  {Sobczyk}},\ }\href {\doibase 10.1103/PhysRevD.79.079903,
  10.1103/PhysRevD.77.053001} {\bibfield  {journal} {\bibinfo  {journal} {Phys.
  Rev.}\ }\textbf {\bibinfo {volume} {D77}},\ \bibinfo {pages} {053001}
  (\bibinfo {year} {2008})},\ \bibinfo {note} {[Erratum: Phys.
  Rev.D79,079903(2009)]},\ \Eprint {http://arxiv.org/abs/0707.3561}
  {arXiv:0707.3561 [hep-ph]} \BibitemShut {NoStop}%
\bibitem [{\citenamefont {{Zmuda J.}}()}]{Zmuda_thesis}%
  \BibitemOpen
  \bibfield  {author} {\bibinfo {author} {\bibnamefont {{Zmuda J.}}},\ }\emph
  {\bibinfo {title} {{Consistent Many-Body Models of Lepton-Nucleus Scattering
  in the Energy Range Between 500 and 1200 MeV}}},\ \href@noop {} {Ph.D.
  thesis}\BibitemShut {NoStop}%
\bibitem [{\citenamefont {de~Swart}\ \emph {et~al.}(1997)\citenamefont
  {de~Swart}, \citenamefont {Rentmeester},\ and\ \citenamefont
  {Timmermans}}]{deSwart:1997ep}%
  \BibitemOpen
  \bibfield  {author} {\bibinfo {author} {\bibfnamefont {J.~J.}\ \bibnamefont
  {de~Swart}}, \bibinfo {author} {\bibfnamefont {M.~C.~M.}\ \bibnamefont
  {Rentmeester}}, \ and\ \bibinfo {author} {\bibfnamefont {R.~G.~E.}\
  \bibnamefont {Timmermans}},\ }\bibfield  {booktitle} {\emph {\bibinfo
  {booktitle} {{Meson nucleon physics and the structure of the nucleon.
  Proceedings, 7th International Symposium, MENU'97, Vancouver, Canada, July
  28-August 1, 1997}}},\ }\href@noop {} {\bibfield  {journal} {\bibinfo
  {journal} {PiN Newslett.}\ }\textbf {\bibinfo {volume} {13}},\ \bibinfo
  {pages} {96} (\bibinfo {year} {1997})},\ \Eprint
  {http://arxiv.org/abs/nucl-th/9802084} {arXiv:nucl-th/9802084 [nucl-th]}
  \BibitemShut {NoStop}%
\bibitem [{\citenamefont {Hernandez}\ \emph {et~al.}(2010)\citenamefont
  {Hernandez}, \citenamefont {Nieves}, \citenamefont {Valverde},\ and\
  \citenamefont {Vicente~Vacas}}]{Hernandez:2010bx}%
  \BibitemOpen
  \bibfield  {author} {\bibinfo {author} {\bibfnamefont {E.}~\bibnamefont
  {Hernandez}}, \bibinfo {author} {\bibfnamefont {J.}~\bibnamefont {Nieves}},
  \bibinfo {author} {\bibfnamefont {M.}~\bibnamefont {Valverde}}, \ and\
  \bibinfo {author} {\bibfnamefont {M.~J.}\ \bibnamefont {Vicente~Vacas}},\
  }\href {\doibase 10.1103/PhysRevD.81.085046} {\bibfield  {journal} {\bibinfo
  {journal} {Phys. Rev.}\ }\textbf {\bibinfo {volume} {D81}},\ \bibinfo {pages}
  {085046} (\bibinfo {year} {2010})},\ \Eprint {http://arxiv.org/abs/1001.4416}
  {arXiv:1001.4416 [hep-ph]} \BibitemShut {NoStop}%
\bibitem [{\citenamefont {Wu}\ \emph {et~al.}(2015)\citenamefont {Wu},
  \citenamefont {Sato},\ and\ \citenamefont {Lee}}]{Wu:2014rga}%
  \BibitemOpen
  \bibfield  {author} {\bibinfo {author} {\bibfnamefont {J.-J.}\ \bibnamefont
  {Wu}}, \bibinfo {author} {\bibfnamefont {T.}~\bibnamefont {Sato}}, \ and\
  \bibinfo {author} {\bibfnamefont {T.~S.~H.}\ \bibnamefont {Lee}},\ }\href
  {\doibase 10.1103/PhysRevC.91.035203} {\bibfield  {journal} {\bibinfo
  {journal} {Phys. Rev.}\ }\textbf {\bibinfo {volume} {C91}},\ \bibinfo {pages}
  {035203} (\bibinfo {year} {2015})},\ \Eprint {http://arxiv.org/abs/1412.2415}
  {arXiv:1412.2415 [nucl-th]} \BibitemShut {NoStop}%
\bibitem [{\citenamefont {Abe}\ \emph {et~al.}(2013)\citenamefont {Abe} \emph
  {et~al.}}]{Abe:2012av}%
  \BibitemOpen
  \bibfield  {author} {\bibinfo {author} {\bibfnamefont {K.}~\bibnamefont
  {Abe}} \emph {et~al.} (\bibinfo {collaboration} {T2K}),\ }\href {\doibase
  10.1103/PhysRevD.87.012001, 10.1103/PhysRevD.87.019902} {\bibfield  {journal}
  {\bibinfo  {journal} {Phys. Rev.}\ }\textbf {\bibinfo {volume} {D87}},\
  \bibinfo {pages} {012001} (\bibinfo {year} {2013})},\ \bibinfo {note}
  {[Addendum: Phys. Rev.D87,no.1,019902(2013)]},\ \Eprint
  {http://arxiv.org/abs/1211.0469} {arXiv:1211.0469 [hep-ex]} \BibitemShut
  {NoStop}%
\bibitem [{\citenamefont {Vermaseren}(2000)}]{Vermaseren:2000nd}%
  \BibitemOpen
  \bibfield  {author} {\bibinfo {author} {\bibfnamefont {J.~A.~M.}\
  \bibnamefont {Vermaseren}},\ }\href@noop {} {\  (\bibinfo {year} {2000})},\
  \Eprint {http://arxiv.org/abs/math-ph/0010025} {arXiv:math-ph/0010025
  [math-ph]} \BibitemShut {NoStop}%
\bibitem [{\citenamefont {Dufner}\ and\ \citenamefont
  {Tsai}(1968)}]{Dufner:1967yj}%
  \BibitemOpen
  \bibfield  {author} {\bibinfo {author} {\bibfnamefont {A.~J.}\ \bibnamefont
  {Dufner}}\ and\ \bibinfo {author} {\bibfnamefont {Y.-S.}\ \bibnamefont
  {Tsai}},\ }\href {\doibase 10.1103/PhysRev.168.1801} {\bibfield  {journal}
  {\bibinfo  {journal} {Phys. Rev.}\ }\textbf {\bibinfo {volume} {168}},\
  \bibinfo {pages} {1801} (\bibinfo {year} {1968})}\BibitemShut {NoStop}%
\end{thebibliography}%

\end{document}